\documentstyle[onecolumn]{mn}

\newcounter{parentequation}
\def\beglet{
  \addtocounter{equation}{1}%
  \setcounter{parentequation}{\value{equation}}%
  \setcounter{equation}{0}%
  \def\theequation{\arabic{parentequation}\alph{equation}}%
  \ignorespaces
}
\def\endlet{
  \setcounter{equation}{\value{parentequation}}%
  \def\theequation{\arabic{equation}}%
}

\def\ltsima{$\; \buildrel < \over \sim \;$}
\def\gtsima{$\; \buildrel > \over \sim \;$}
\def\simlt{\lower.5ex\hbox{\ltsima}}
\def\simgt{\lower.5ex\hbox{\gtsima}}

\def\etal{{\it et al.}\rm}
\def\etals{{\it et al. }\rm}
\def\mk2{\mu {\rm K}^2}

\begin{document}

\title[Power Spectrum Estimation]
{Myths and Truths Concerning Estimation of Power Spectra: The Case
for a Hybrid Estimator}

 \author[G. Efstathiou]{G. Efstathiou\\
Institute of Astronomy, Madingley Road, Cambridge, CB3 OHA.}

\maketitle

\begin{abstract}
It is widely believed that maximum likelihood estimators must be used
to provide optimal estimates of power spectra. Since such estimators
require the inversion and multiplication of $N_d\times N_d$ matrices,
where $N_d$ is the size of the data vector, maximum likelihood
estimators require at least of order $N_d^3$ operations and become
computationally prohibitive for $N_d$ greater than a few tens of
thousands. Because of this, a large and inhomogeneous literature
exists on approximate methods of power spectrum estimation. These
range from manifestly sub-optimal, but computationally fast methods,
to near optimal but computationally expensive methods. Furthermore,
much of this literature concentrates on the power spectrum estimates
rather than the equally important problem of deriving an accurate
covariance matrix. In this paper, I consider the problem of estimating
the power spectrum of cosmic microwave background (CMB) anisotropies
from large data sets. Various analytic results on power spectrum
estimators are derived, or collated from the literature, and tested
against numerical simulations. An unbiased hybrid estimator is
proposed that combines a maximum likelihood estimator at low
multipoles and pseudo-$C_\ell$ estimates at high multipoles.  The
hybrid estimator is computationally fast ({\it i.e.}  it can be run on
a laptop computer for Planck sized data sets), nearly optimal over the
full range of multipoles, and returns an accurate and nearly diagonal
covariance matrix for realistic experimental configurations (provided
certain conditions on the noise properties of the experiment are
satisfied). It is argued that, in practice, computationally expensive
methods that approximate the ${\cal O}(N_d^3)$ maximum likelihood
solution are unlikely to improve on the hybrid estimator, and may
actually perform worse. The results presented here can be generalised
to CMB polarization and to power spectrum estimation using other types
of data, such as galaxy clustering and weak gravitational lensing.

\vskip 0.1 truein


\vskip 0.3 truein

\end{abstract}

\section{Introduction}

The estimation of power spectra from various types of data is becoming
increasingly important in cosmology. There are three main reasons for
this. Firstly, there has been an explosion in the size and quality of
cosmological data sets, for example, the detailed maps of the CMB sky
from WMAP\footnote{WMAP: Wilkinson Microwave Anisotropy Probe; 2dFGRS:
Two Degree Field Galaxy Redshit Survey; SDSS: Sloan Digital Sky
Survey.} (Bennett \etals 2003), the 2dFGRS (Colless \etals 2001) 
and SDSS galaxy redshift survey (Zehavi \etals 2002; Tegmark \etals 2004). With such large data
sets it is now possible to measure power spectra accurately over a
wide range of angular and spatial scales. Secondly, the power spectrum
is a simple two-point statistic and so it is natural that more effort
has been devoted to optimal methods of power spectrum estimation
rather than to optimal estimators for higher order
statistics. Thirdly, in most realisations of the inflationary
scenario, the fluctuations generated during the inflationary phase are
predicted to be Gaussian (see {\it e.g.} Liddle and Lyth, 2000, for a
review). If the fluctuations are Gaussian, and as long as they remain
linear, all of the information pertaining to the fluctuations is
contained in the power spectrum. In this case, estimation of the power
spectrum can be viewed as a form of lossless data compression, reducing,
for example, $10^7$ measurements of the CMB temperature differences
$\Delta T_i$ into $2000$ or so values of the angular power spectrum
$C_\ell$ (see {\it e.g.} Tegmark 1997). This form of data compression
is invaluable, since physical parameters of interest (matter
densities, spectral indices {\it etc}) can be estimated from the power
spectrum and its covariance matrix, rather than from the pixel values
themselves.

 One of the earliest applications of power spectrum analysis to
cosmology was Yu and Peebles' (1969) analysis of the angular
distribution of rich clusters of galaxies. These authors applied what
is now (unfortunately) commonly referred to as a `pseudo-$C_\ell$'
estimator (see Section 2). In the rest of this paper, we will discuss
the specific problem of power spectrum estimation of CMB temperature
anisotropies, although many of the results and remarks are applicable
to other types of data, such as galaxy clustering. A comprehensive
analysis of pseudo-$C_\ell$ estimators was given by Peebles (1973) and
an application to the angular clustering of galaxies can be found in
Peebles and Hauser (1974). Although it was well known that
pseudo-$C_\ell$ estimators are sub-optimal, little work was done to
develop more optimal power spectrum estimators until recently (the
paper by Feldman, Kaiser and Peacock, 1994, on power spectrum analysis
of redshift surveys is a noteable exception). Hamilton (1997a, b),
Tegmark (1997) and Bond, Jaffe and Knox (1998) constructed estimators
that are optimal, in the sense that they find the power spectrum that
maximises the likelihood function given the data. However, there is a
fundamental problem with these maximum likelihood estimators. Given a
data vector of length $N_d$, optimal estimators require the inversion
of $N_d \times N_d$ matrices. Since a matrix inversion requires 
${\cal O}(N_d^3)$ operations, brute force application of maximum likelihood estimators
becomes impractical for $N_d$ greater than a few tens of
thousands\footnote{Perhaps $100,000$ if one has access to a
supercomputer.} (see {\it e.g.} Borrill 1999a,b). As a result, a large
number of papers have appeared in the last few years that discuss
various solutions to the problem of power spectrum estimation from
large data sets.

 These papers can be grouped, roughly, into the following categories:

\smallskip

\noindent
{\it (i) Pseudo-$C_\ell$ estimators:}  These are straightforward
variants of the methods described by Peebles (1973).  A discussion of
a pseudo-$C_\ell$ estimator (hereafter PCL) applied to CMB experiments
is given by Wandelt, Hivon and G\'orski (2001) and Hivon \etals
(2002). Variants designed to recover the true shape of the power
spectrum from apodised regions of the sky are discussed by Hansen,
G\'orski and Hivon (2002) and Hansen and G\'orski (2003). These estimators
can be evaluated using fast spherical transforms\footnote{Or Fast
Fourier Transforms in three dimensions.}  ({\it e.g.}  Muciaccia,
Natoli and Vittorio 1997) for which the number of operations scale as
$N_d^{3/2}$. Even for data vectors of the length expected from the
Planck mission (Bersannelli \etals 1996), $N_d \sim 10^7$, it is
possible to evaluate a PCL estimator using a laptop computer (see {\it
e.g.} Balbi \etals 2002). A PCL estimator was used in the analysis of
the WMAP data (Hinshaw \etals 2003)
  
  A related class of sub-optimal estimators use fast evaluation of the
two-point correlation function, which can then be transformed to give
an estimate of the power spectrum. Methods of this type, designed for
the analysis of CMB anisotropies are described by Szapudi \etals
(2001a) and by Szapudi, Prunet and Colombi (2001b). A generalisation of
these methods to the analysis of CMB polarization is discussed by Chon
\etals (2003). We group this type of estimator along with the PCL
methods described in the previous paragraph because
they are almost mathematically equivalent.

\smallskip

\noindent
{\it (ii) Maximum likelihood estimators:}
Given a data vector $x_i$ of length $N_d$, and assuming that the $x_i$
are Gaussian distributed, we can define a likelihood function
\begin{equation}
{\cal L} (C_\ell \vert {\bf x}) = {
{ \rm exp} \left ( -{1 \over 2} x^T C^{-1} x \right ) \over  (2 \pi)^{N_d/2} 
\vert C \vert^{1/2}},  \label{I1}
\end{equation}
where $C_\ell$ is the power spectrum to be estimated. The covariance 
matrix $C$ in (\ref{I1}) is 
\begin{equation}
 C_{ij} = \langle x_i x_j \rangle = S_{ij}(C_\ell) + N_{ij}, \label{I2}
\end{equation}
where $S$ is the signal covariance matrix and $N$ is the noise matrix.
A maximum likelihood estimator finds the $C_\ell$, or an approximation to the
$C_\ell$, that maximises the likelihood (\ref{I1}).

Two types of maximum likelihood (herafter ML) estimator have been discussed widely.
In the first, which we will refer to as NRML, the power spectrum that
maximises the likelihood function is found iteratively using a
Newton-Raphson algorithm (Bond, Jaffe and Knox 1998; Ruhl \etals
2003).  In the second, which we will refer to as QML, a quadratic
estimator is defined based on an assumed form for $C_\ell$ which
is equivalent to a maximum likelihood solution if the guess for $C_\ell$
is close to the true power spectrum ({\it e.g.} Tegmark
1997).  From equation (1) one can see that both the NRML and QML
estimators require the inverse and determinant of the covariance
matrix $C$ and so require of  ${\cal O}(N_d^3)$ operations.

Various methods have been proposed to speed up the computation of ML
estimators. Oh, Spergel and Hinshaw (1999) use a conjugate gradient
algorithm with a carefully chosen preconditioner to avoid direct
invertion of the matrix (\ref {I2}). Dor\'e, Knox and Peel (2001)
propose an NRML estimator based on hierarchical decomposition of a
large map into sub-maps of various resolutions, though this method has
the disadvantage of a loss of resolution in the power spectrum
estimates.  An iterative multi-grid method is described by Pen (2003)
and a method that uses Monte Carlo Markov Chains is described by
Jewell, Levin and Anderson (2002).

\smallskip

\noindent
{\it (iii) Harmonic and Ring-torus Estimators:}
For a Planck-type scanning strategy, a ML  solution for
the spherical harmonic coefficients $a_{\ell m}$ can be determined
from the Fourier transform of the temperature differences on a set of
rings on the sky (van Leeuwen \etals 2002; Challinor \etals
2002). The data vector $x_i$ in equation (\ref{I1}) is then the set of
harmonic coefficients $a_{\ell m}$.  This type of `harmonic' method
avoids the need to construct pixelised maps of the temperature
differences on the sky and can account for asymmetric beams and
certain types of correlated noise. However, a brute force application
of the harmonic method would require of ${\cal O}(\ell_{\rm max}^6)$
operations, where $\ell_{\rm max}$ is the highest multipole under
consideration, and so is not computationally feasible for high
multipoles. The operation count can be reduced for simple scanning
strategies and by using conjugate gradient techniques, but so far
little work has been done along these lines. A related method is
described by Wandelt and Hansen (2003), who show that for special
scanning geometries the time-ordered data (TOD) can be wrapped onto a ring
torus. In this case, the ML solution for $C_\ell$ can
be found in ${\cal O}(N^2_d)$ operations using the fast spherical
convolution algorithm described by Wandelt and G\'orski (2001). Even if
the scanning geometry of an experiment does not satisfy the precise
conditions for an ${\cal O}(N^2_d)$ solution, Wandelt and Hansen
argue that the ring-torus method can be useful in determining a 
preconditioner to speed up the computation of the exact ML
solution.

Given the large number of methods described in the previous
paragraphs, what is the best way to estimate a power spectrum from a
large data set, including realistic sources of errors?  In the view of
this author, the work summarised above suggests two approaches:
\footnote{After this paper was submitted for publication, Wandelt
Larson and Lakshminarayanan (2003) have described an ingeneous method
which does not fit into either of the two approaches described
here. Their method, which is related to the method of Jewell \etals
(2002), can recover the joint posterior distribution of the CMB power
spectrum and signal directly from the TOD using Gibbs resampling . The
method is potentially very powerful because it can account for complex
scanning patterns and noise properties and furthermore scales as
$N_d^{3/2}$, though with a large prefactor. Application of this technique
to Planck size data sets poses a challenging, but perhaps not insurmountable,
computational problem.}

\noindent
${\bf (A)}$ Application of a computationally expensive ML method that
can account for various sources of error, such as correlated receiver
noise, beam asymmetries {\it etc}. For a large ($N_d \simgt 10^6$)
data set, the noise matrix $N$ in equation (\ref{I2}) is neither
calculable nor storeable, and so if correlated noise is to be taken
into account the method would have to be based on (and exploit
symmetries in) the TOD.

\noindent
${\bf (B)}$ Application of a fast PCL-based method which can be used to
derive an analytic covariance matrix under certain simplifying  assumptions ({\it e.g.}
diagonal noise matrix, symmetric beams). Real-world complications can then
be treated as perturbations to the covariance matrix, which can be estimated
by running a large number of Monte Carlo simulations.

The point of view adopted in this paper is that approach (B) is preferable
to approach (A) and that there has been an over-emphasis in the literature
on computationally expensive maximum likelihood methods. There are several
reasons for taking this viewpoint:

\noindent
{\it (a)} Does a maximum likelihood estimator lead to a significant
reduction in the error bars compared to using a PCL estimator? In some
limits it is possible to prove that a PCL is statistically equivalent
to a ML estimator (Sections 3 and 5). Furthermore, for realistic
experiments there may be `irreducible' errors on the power spectrum
estimates that dominate over any minor differences between
estimators. For example, in the WMAP experiment beam calibration
uncertainties are the dominant source of error in the range $100
\simlt \ell \simlt 600$ and residual Galactic emission is the dominant
source of error at multipoles $\ell \simlt 10$ (Hinshaw \etals 2003).

\noindent
{\it (b)} As discussed above, for many purposes, a power spectrum is
simply a convenient form of data compression to enable fast estimation
of a small number of physical parameters. But to determine the
physical parameters we must be able to write down a likelihood
function, which requires at least an accurate covariance matrix for
the power spectrum estimates and preferably an accurate model for the shape of the
likelihood function (see {\it e.g.} Bond, Jaffe and Knox, 2000). There
is little point in applying a computationally expensive method of
power spectrum estimation, which may only produce a marginal
improvement in the estimates themselves, if the method returns a poor
estimate of the covariance matrix. {\it The estimation of an accurate covariance
matrix is of comparable importance to the estimation of the power spectrum}.

\noindent
{\it (c)} Following on from point (b), we must be able to demonstrate
that any power spectrum estimation method produces negligible, or
acceptable, bias on the {\it physical} parameters of interest. It is
not enough simply to demonstrate that a method recovers the power
spectrum with a bias that is small  compared to the variance of the estimator,
since small correlated errors can cause a significant bias in the
estimation of physical parameters (see Efstathiou and Bond, 1999,
Section 5).

\noindent
{\it (d)} A fast PCL method is much more flexible than a
computationally expensive ML method. For example, it may be feasible
to implement an ML method by exploiting symmetries appropriate to a 
particular scanning pattern. However, it may then be difficult to
adapt the method to another scanning strategy. A realistic experiment
will contain intentional and unforseen gaps in the TOD which might be
difficult to handle in an ML method. In a ML method based on TOD, it
may be difficult to deal with foreground component separation and
Galactic sky cuts, or the method may impose significant restrictions on the
algorithms used for component separation. These types of complexity can
be incorporated easily into a PCL method, using Monte Carlo
simulations if necessary.

The point of view adopted here is not new. It has been argued before
by Hivon \etals (2002) and is implicit in the analysis of the WMAP
experiment (Bennett \etals 2003; Hinshaw \etals 2003). The purpose of
this paper is to collate and derive various results that provide
additional, and in the view of this author compelling, support for
this viewpoint. In the first part of this paper (Sections 2-4) we
consider the idealised case of a CMB experiment free of instrumental
noise. PCL estimators are discussed in Section 2 and expressions for
the covariance matrix are derived and tested against simulations.  A
QML estimator is discussed in Section 3 and is shown to be
statistically equivalent to a PCL estimator at high multipoles. An
unbiased hybrid estimator is proposed in Section 4 that combines a QML
estimator at low multipoles and a PCL estimator at high multipoles. It
is argued that this hybrid estimator, which can be evaluated on a
laptop computer and returns an accurate estimate of the covariance
matrix, is virtually indistinguishable from a full ML solution. Inclusion of
instrumental noise is discussed in Section 5 and the hybrid estimator
is generalized to include more than one PCL estimate with different
pixel weights.  In the presence of instrumental noise, it is argued
that a fast hybrid estimator can be defined that is close to optimal
for all multipoles. This is demonstrated by applying a hybrid
estimator to a high resolution simulation of the CMB sky ($0.2^\circ$
pixels), including a Planck-type scanning pattern. The conclusions
are presented in Section 6, together with a discussion of possible
limitations of the hybrid estimator and areas where further work
is required.

\section{Estimation using Pseudo-$C_\ell$}

\subsection{The Pseudo-$C_\ell$ estimator}

We first review some basic results and establish some notation.
The spherical harmonic transform of a  map $\Delta T_i$ is defined as
\begin{equation}
 \tilde a_{\ell m} = \sum_i \Delta T_i w_i 
\Omega_i Y_{\ell m}({\bf \theta}_i),    \label{PCL1}
\end{equation}
where $\Omega_i$ is the area of pixel $i$ and $w_i$ is an arbitrary
weight function with spherical transform
\begin{equation}
 \tilde w_{\ell m} = \sum_i w_i
\Omega_i Y_{\ell m}({\bf \theta}_i).     \label{PCL2}
\end{equation}
The $\tilde a_{\ell m}$ coefficients are related to the true $a_{\ell m}$
coefficients on the uncut sky by a coupling matrix $K$,
\beglet
\begin{equation}
 \tilde a_{\ell m} = \sum_{\ell^\prime m^\prime} a_{\ell^\prime m^\prime}
K_{{\ell m}{\ell^\prime m^\prime}}, \label{PCL3}
\end{equation}
where
\begin{equation}
  K_{{\ell_1 m_1}{\ell_2 m_2}} =
\sum_{\ell_3 m_3} \tilde  w_{\ell_3 m_3}
\left ( {(2 \ell_1 + 1) (2 \ell_2 + 1) (2 \ell_3 + 1) \over 4 \pi} \right )^{1/2} (-1)^{m_2} (-1)^{m_3} 
{\left ( \begin{array}{ccc}
        \ell_1 & \ell_2 & \ell_3  \\
        0  & 0 & 0
       \end{array} \right )
 \left ( \begin{array}{ccc}
        \ell_1 & \ell_2 & \ell_3  \\
         m_1  & -m_2 & -m_3
       \end{array} \right ), }  \label{PCL4}
\end{equation}
\endlet
(see {\it e.g.} Hivon \etals 2002).
An alternative expression for $K$, in terms of a sum over pixels
is
\begin{equation}
 K_{{\ell_1 m_1}{\ell_2 m_2}} = \sum_i w_i \Omega_i Y_{\ell_1 m_1} (
{\bf \theta}_i) Y^*_{\ell_2 m_2} ( {\bf \theta}_i) . \label{PCL5}
\end{equation} 

A PCL estimator is constructed from the sum
\begin{equation}
   \tilde C^p_\ell = {1 \over (2 \ell + 1)} \sum_m \vert \tilde a_{\ell m} \vert ^2.  \label{PCL6}
\end{equation}
Since the weight function $w_i$ is unspecified, equation (\ref{PCL6})
defines a {\it set} of PCL estimators. As is well known (Peebles 1973;
Hivon \etals 2002), the expectation value of (\ref{PCL6}) is related
to the true power spectrum $C_\ell$ by a convolution
\begin{equation}
\langle  \tilde C^p_{\ell } \rangle = \sum_{\ell^\prime } C_{\ell^\prime} M_{\ell \ell^\prime}, \label{PCL7}
\end{equation}
where the coupling matrix can be expressed in terms of $3j$ symbols as
\begin{equation}
  M_{\ell_1 \ell_2}  =  (2 \ell_2 + 1)
\sum_{\ell_3 }  {(2 \ell_3 + 1) \over 4 \pi}\tilde W _{\ell_3}
{\left ( \begin{array}{ccc}
        \ell_1 & \ell_2 & \ell_3  \\
        0  & 0 & 0
       \end{array} \right )^2} \label{PCL8}
\end{equation}
and  $\tilde W_\ell$ (the `window function') is the power spectrum of the weighting function
$\tilde w_{\ell m}$, defined in an analogous way to equation (\ref{PCL6}). The summation in 
equation (\ref{PCL8}) will appear several times in this paper, so to simplify the
notation we define
\begin{equation}
  \Xi(\ell_1, \ell_2, \tilde W)  =  
\sum_{\ell_3 }  {(2 \ell_3 + 1) \over 4 \pi}\tilde W _{\ell_3}
{\left ( \begin{array}{ccc}
        \ell_1 & \ell_2 & \ell_3  \\
        0  & 0 & 0
       \end{array} \right )^2}. \label{PCL8a}
\end{equation}
In addition, for some purposes it is useful to define the explicitly symmetric 
coupling matrix $G$,
\begin{equation}
 G_{\ell_1 \ell_2}  = {1 \over (2 \ell_2 + 1) } M_{\ell_1 \ell_2}. \label{PCL9}
\end{equation}

If  the coupling matrix $M_{\ell_1 \ell_2}$ is invertible,  unbiased 
(deconvolved) estimates $\hat C^p_{\ell}$ of the true power spectrum can be reconstructed
via 
\begin{equation}
 \hat C^p_\ell = M^{-1}_{\ell \ell^\prime} \tilde C^p_{\ell^\prime}. \label{PCL10}
\end{equation} 
Evidently, provided $M_{\ell_1 \ell_2}$ is invertible, 
unbiased estimates of $C_\ell$ can be recovered independently of the weights $w_i$. The
only effect of the weights is to change the covariance matrix and
hence we should seek weights that minimise the errors.

\begin{figure*}

\vskip 3.8 truein

\includegraphics{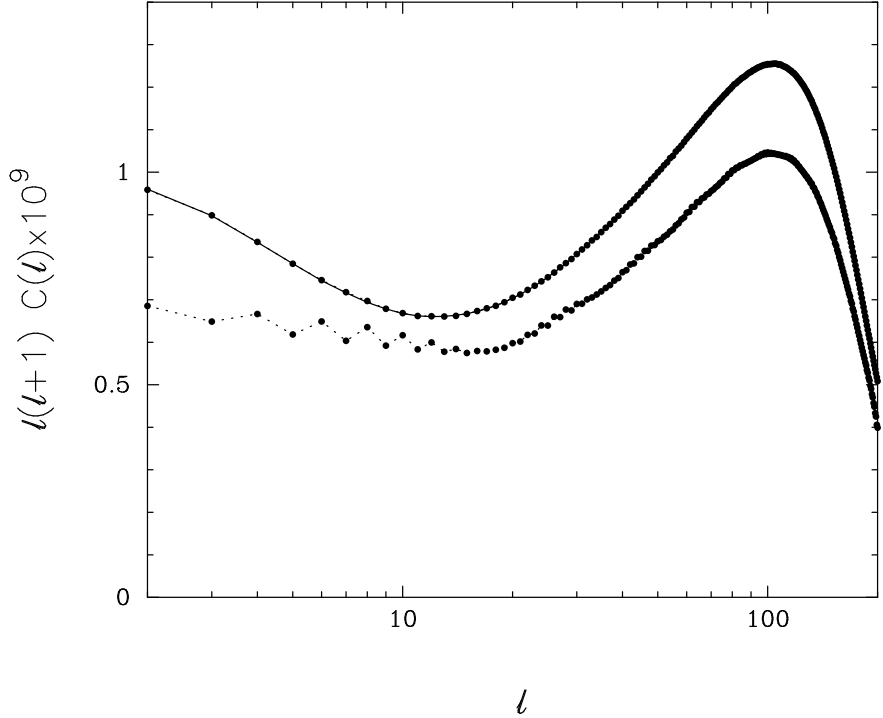}

\caption
{The points joined by the dotted line show the average values of the PCL estimates, 
$\tilde C^p_\ell$ from $10^5$ simulations using equal weights per pixel outside a
Galactic cut of $\pm 10^\circ$.
The points joined by the solid line show average values of
the deconvolved estimates $\hat C^p_\ell$. The solid line shows the input
theoretical power spectrum and the dotted line shows this spectrum convolved with the
coupling matrix $M$ (equation \ref{PCL7}).}

\label{figure1}

\end{figure*}

In the analysis of CMB experiments, a region close to the Galactic
plane is often excised from a map to reduce contamination of the
primordial CMB signal from Galactic emission. If the Galactic cut is
small enough, then the coupling matrix $M$ will be invertible. As a
rough rule of thumb, the matrix $M$ will be invertible if the
two-point correlation function $C(\theta)$ can be determined on all
angular scales from the data within the uncut sky (Mortlock, Challinor
and Hobson 2002). In practice this means that quite large sky cuts (up
to $\sim 30^\circ$ above and below the Galactic plane) can be imposed
on the data before the matrix $M$ becomes singular (see also
Efstathiou 2004). In contrast, for a finite sky cut, the matrix
$K_{\ell m \ell m^\prime}$ defined in equation (\ref {PCL3}) will be
singular, since some linear combinations of the $a_{\ell m}$ will
define modes that lie within the cut sky. Obviously, these `null
modes' cannot be recovered from the observed $\tilde a_{\ell m}$.

Figure \ref{figure1} shows an example of PCL estimation applied to simulated CMB
skies. Throughout this paper we will adopt the CMB power spectrum of
the concordance $\Lambda$CDM model favoured by WMAP (Spergel \etals
2003).  The exact parameters adopted are those of the fiducial
$\Lambda$CDM model defined in Section 2 of Efstathiou (2003). For the
tests shown in Figure 1, $10^5$ simulations were generated with $\ell_{\rm
max}=200$, a Gaussian beam with FWHM of $\theta_s = 1.0^\circ$ and a pixel size
of $0.5^\circ$. The simulations described in this paper use an `igloo'
pixelization scheme (identical pixels on lines of constant latitude,
see Figure 7 below and Crittenden and Turok 1998) and fast spherical
transforms based on software developed by the author for the Planck
Phase A study (Bersanelli \etals 1996).  The simulations are
noise-free and a Galactic cut of $\pm 10^\circ$ was imposed.  Unless
otherwise stated, beam functions will not be written explicitly in
equations, thus $C_\ell$ will sometimes mean $C_\ell b_\ell^2$. Since
the symmetric beams are used in all of the simulations in this paper
including beams simply involves multiplying theoretical power spectra,
and dividing estimated power spectra, by the beam function $b_\ell^2$.
In all of the figures in this paper, the power spectra are divided
by $T_0^2$, so that they are dimensionless, where $T_0=2.275$ is the
mean CMB temperature (Mather \etals 1999).

The points connected by the dotted line in Figure \ref{figure1} show
the average, over $10^5$ simulations, of the PCL estimates $\tilde
C^p_{\ell}$ computed from equation (\ref{PCL6}) using equal weight per
pixel in the region outside the Galactic cut. The points joined by the
solid line show the average of the deconvolved estimates $\hat
C^p_\ell$ (equation \ref{PCL10}). Neither of these estimates has been
corrected for the Gaussian beam. These results are shown to
demonstrate that it is possible to run large numbers of simulations of
the PCL method without detecting any significant bias.  Issues related
to the finite pixel sizes, and the evaluation of discrete spherical
transforms (Crittenden and Turok 1998; G\'orski \etals 1999;
Doroshkevich \etals 2004), can be kept under control by choosing  pixels 
that are small enough and so will not be discussed further in this paper.

We end this sub-section with some remarks concerning PCL estimation
when the coupling matrix $M$ is singular. If the power spectrum is to
be used to estimate physical parameters, then clearly there is no need
to construct a deconvolved PCL estimate from equation
(\ref{PCL10}). The PCL estimate (\ref{PCL6}) and its covariance matrix
(Section 2.2) can be used to construct a likelihood function without
any loss of information. If the matrix $M$ is singular, then any
attempt to compute an approximation to the deconvolved estimate $\hat
C^p_\ell$ will entail loss of information and will provide a poorer
approximation to the likelihood function than the convolved estimates
$\tilde C^p_\ell$. Nevertheless, there may be `cosmetic' reasons for
wanting an approximation to $\hat C^p_{\ell}$, for example, to compare
different experiments on the same plot. In this case, one can solve
for band-power averages of $ \ell(\ell +1) \hat C^p_\ell$ over ranges
in $\ell$ comparable to the width of the window function $\tilde W_\ell$ (see
{\it e.g.} Bond \etals 1998). Alternatively, one can introduce an
invertible regularising matrix $R$ and compute
\begin{equation}
R  \hat C^p_\ell = (R^{-1} M)^{-1} \tilde C^p \label{PCL11}
\end{equation} 
for each experiment. If $R$ is chosen to have a width comparable to the window functions
$\tilde W_\ell$ of the two experiments, the product $R^{-1}M$ should be invertible.

\begin{figure*}
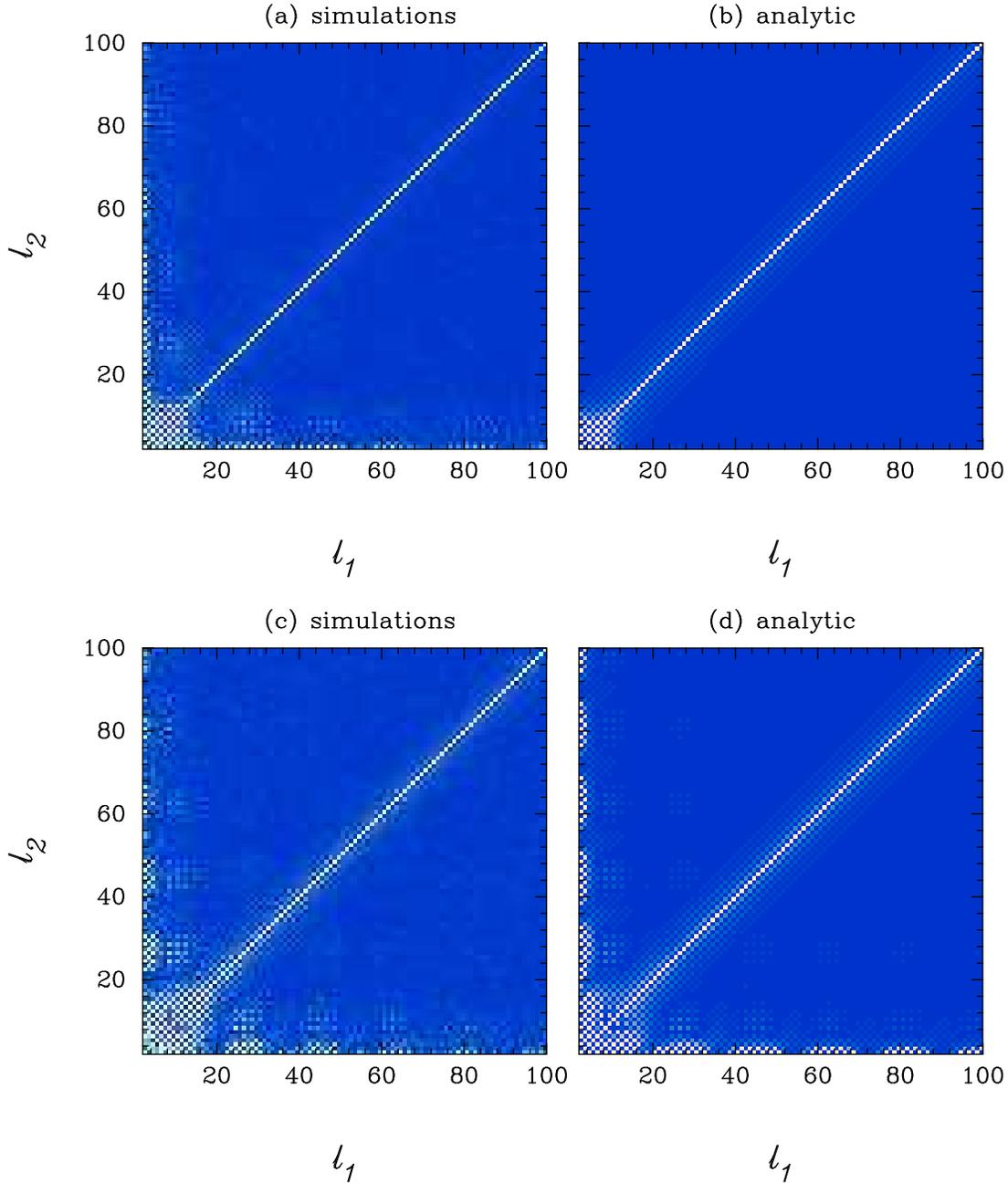


\vskip 6.8 truein

\includegraphics{pgcov2a.ps}
\includegraphics{pgcov2b.ps}

\caption
{The upper panels, (a) and (b), show the covariance matrices for the PCL estimates
$\tilde C^p_\ell$ and the lower panels, (c) and (d), show the covariance matrices
for the deconvolved PCL estimates $\hat C^p_\ell$. Plots to the left
show the covariance matrices averaged over the $10^5$ simulations
described in Section 2.1; plots to the right  show the
show the analytic approximations of equations (\ref{V3}) and (\ref{V4}). The covariance
matrices have been multiplied by $\ell_1^2\ell_2^2$ so that the diagonal components
have roughly equal amplitude and to amplify the off-diagonal elements.}

\label{figure2}

\end{figure*}

\subsection{Covariance matrix of the pseudo-$C_\ell$ estimator}

Starting from the defininition of the PCL estimator (equations \ref{PCL1} and
\ref{PCL6}) it is straightforward to show that the covariance matrix is 
given by
\begin{equation}
 \langle \Delta \tilde C^p_\ell 
\Delta \tilde C^p_{\ell^\prime} \rangle = 
 {2 \over (2 \ell + 1) (2 \ell^\prime + 1) } 
\sum_{m m^\prime} \sum_{\ell_1 m_1} \sum_{\ell_2 m_2}
C_{\ell_1} C_{\ell_2} K_{\ell m \ell_1 m_1} K^*_{\ell^\prime m^\prime \ell_1 m_1} K^*_{\ell m \ell_2 m_2} K_{\ell^\prime m^\prime \ell_2 m_2}. \label{V1}
\end{equation}
As it stands, equation (\ref{V1}) is not useful, but it can be
simplified for high multipoles (greater than the width of the window function
$\tilde W_\ell$). If the Galactic cut is narrow, then we
can replace $C_{\ell_1}$ and $C_{\ell_2}$ with $C_\ell$ and
$C_{\ell^\prime}$ and then apply the completeness relation for
spherical harmonics ({\it e.g.} Varshalovich, Moskalev and Khersonskii
1988). This gives
\beglet
\begin{equation}
 \langle \Delta \tilde C^p_\ell 
\Delta \tilde C^p_{\ell^\prime} \rangle =  \tilde V_{\ell \ell^\prime}
\approx  2 C_\ell C_{\ell^\prime}
\Xi (\ell, \ell^\prime, \tilde W^{(2)}), \label{V2}
\end{equation} 
where $\tilde W^{(2)}_{\ell}$ is the power spectrum of the square of the weight 
function $w_i$,
\begin{equation}
 \tilde W^{(2)}_\ell = {1 \over (2 \ell + 1)} \vert \tilde
w^{(2)}_{\ell m} \vert^2, \qquad \tilde w^{(2)}_{\ell m} = \sum_i w^2_i \Omega_i Y_{\ell m} (\theta_i). \label{V2b}
\end{equation}
\endlet
 If $w_i$ is a simple mask with values of $1$ or $0$,
then $w^2_i \equiv w_i$, $\tilde W^{(2)}_{\ell} \equiv \tilde W_{\ell}$
and so the covariance matrix can be written in terms of the symmetric
coupling matrix $G_{\ell_1 \ell_2}$ defined in equation (\ref{PCL9})
\begin{equation}
 \langle \Delta \tilde C^p_\ell 
\Delta \tilde C^p_{\ell^\prime} \rangle 
\approx 2 C_\ell C_{\ell^\prime}
      G_{\ell \ell^\prime}. \label{V3}
\end{equation} 
The covariance matrix of the deconvolved PCL estimates (equation \ref{PCL10}) is
therefore
\begin{equation}
 \langle \Delta \hat C^p_\ell  \Delta \hat C^p_{\ell^\prime} \rangle = M^{-1} \tilde V (M^{-1})^{T}.
\label{V4}
\end{equation} 
From this equation, and using the general form of $\tilde V$ for arbitrary weights,
we can differentiate the variance  $\langle \Delta \hat C^p_\ell  \Delta \hat 
C^p_{\ell} \rangle$ with respect to $w_i$. After some algebra, we can prove that
for high multipoles, the variance of $\hat C^p_{\ell}$ in a noise-free 
experiment is minimised if we adopt equal weight per pixel. The estimates
shown in Figure 1 are therefore the minimum variance PCL estimates for a noise-free 
experiment. It is possible to define a class of estimators that use a weight
function that maximises the resolution of the power spectrum estimates
from an incomplete sky (Tegmark 1996a, b). However, these require the
inversion of $N_d \times N_d$ matrices and will not be discussed further here.

\begin{figure*}
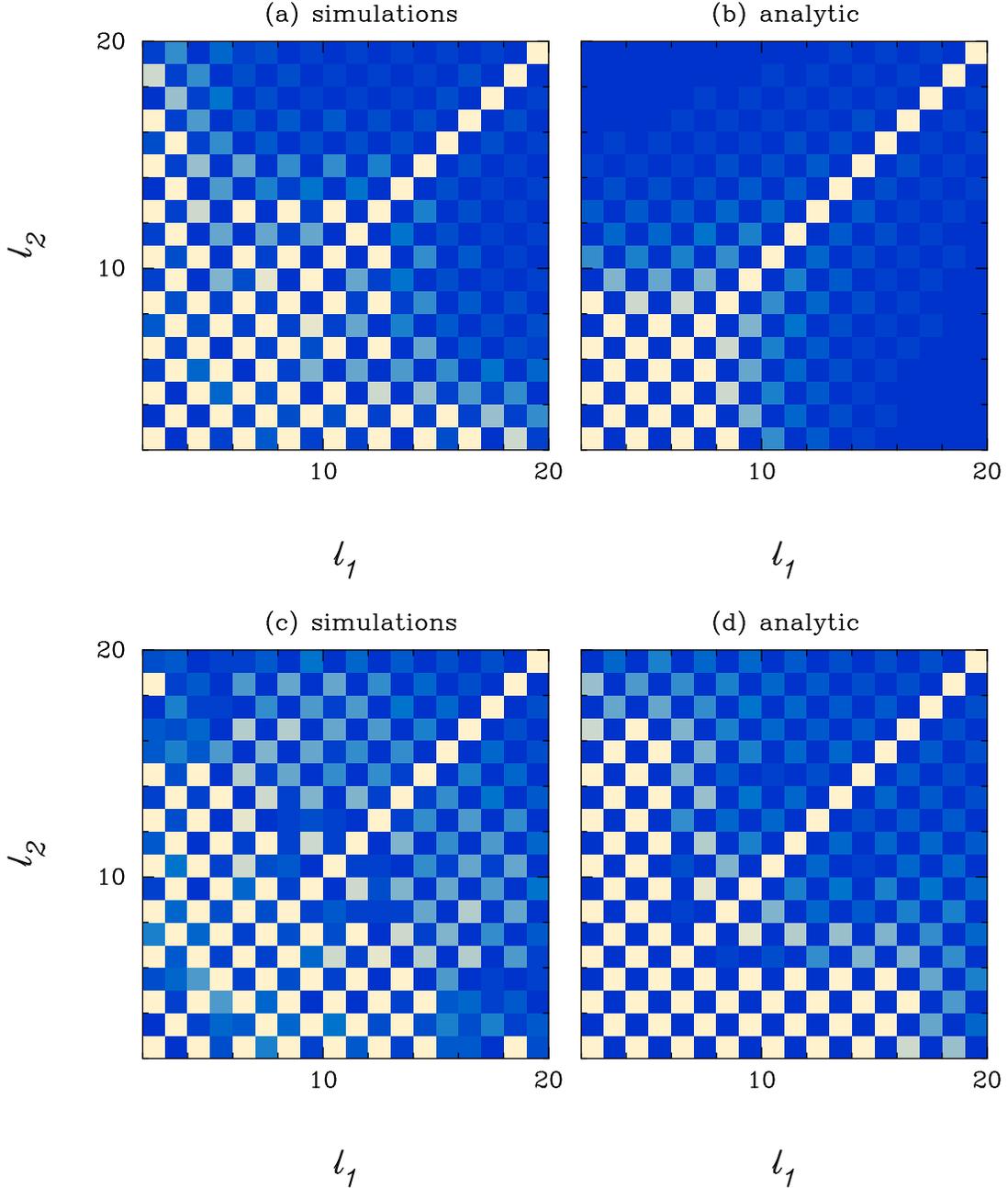


\vskip 6.8 truein

\includegraphics{pgcov30a.ps}
\includegraphics{pgcov30b.ps}

\caption
{As Figure 2, but for $2 \times 10^5$ low resolution simulations 
with pixel size $\theta_c = 5^\circ$ and beam smoothing of $7^\circ$.
The differences between the covariance matrices estimated from the
simulations and the analytic expression (equation \ref{V2})
are now easily seen.}

\label{figure3}

\end{figure*}

Figure 2 compares the covariance matrices determined from the
simulations described in Section 2.1 (left hand panels) with the
analytic expressions of equations (\ref{V2}) and (\ref{V4}). For a
Galactic cut of $\pm 10^\circ$, the analytic approximations are
accurate at $\ell \simgt 20$, where the covariance matrix is very
nearly band-diagonal, but fail at lower multipoles. At $\ell \simlt
20$, even the diagonal components of the analytic covariance matrices
differ (by up to 25\% at $\ell =2$) from the covariance matrices
determined from the simulations (see Figure \ref{figure12} of Section
5.4).  The off-diagonal components differ by much larger factors and
so the analytic approximations at low multipoles are useless for any
quantitative application such as parameter estimation.

This can be seen in Figure 3, which shows a similar comparison to that
shown in Figure 2, but using $2 \times 10^5$ low resolution simulations
with pixel size $\theta_c = 5^\circ$ and Gaussian beam FWHM of $7^\circ$.
However, for such a low resolution map, it is possible to evaluate an 
exact expression for the covariance matrix. This can be done conveniently
by computing
\begin{equation}
 \langle \Delta \tilde C^p_\ell 
\Delta \tilde C^p_{\ell^\prime} \rangle = 
{2  \over (2 \ell + 1) (2 \ell^\prime + 1)} 
\sum_{m m^\prime} \vert B_{\ell m \ell^\prime m^\prime} \vert ^2 \label{V6},
\end{equation} 
where $B_{\ell m \ell^\prime m^\prime}$ is evaluated by summing over pixels
\begin{equation}
 B_{\ell m \ell^\prime m^\prime} = 
\sum_{ij} w_i w_j \Omega_i \Omega_j
C(\theta_{ij}) Y_{\ell m} ({\bf \theta}_i) Y^*_{{\ell^\prime} m^\prime} ({\bf
\theta}_j). \label{V7}
\end{equation} 
In equation (\ref{V7}),  $C(\theta)$ is the temperature autocorrelation function
\begin{equation}
C(\theta_{ij}) = \langle \Delta T_i \Delta T_j \rangle = {1 \over 4 \pi} \sum_{\ell}
(2 \ell + 1) C_\ell P_\ell(\cos \theta). \label{V8}
\end{equation}
 A comparison of the simulations with the exact expression (\ref{V6})
is shown in Figure 4. As expected, the agreement is almost perfect and
any residual differences are simply caused by noise from the finite number of
simulations.

\begin{figure*}
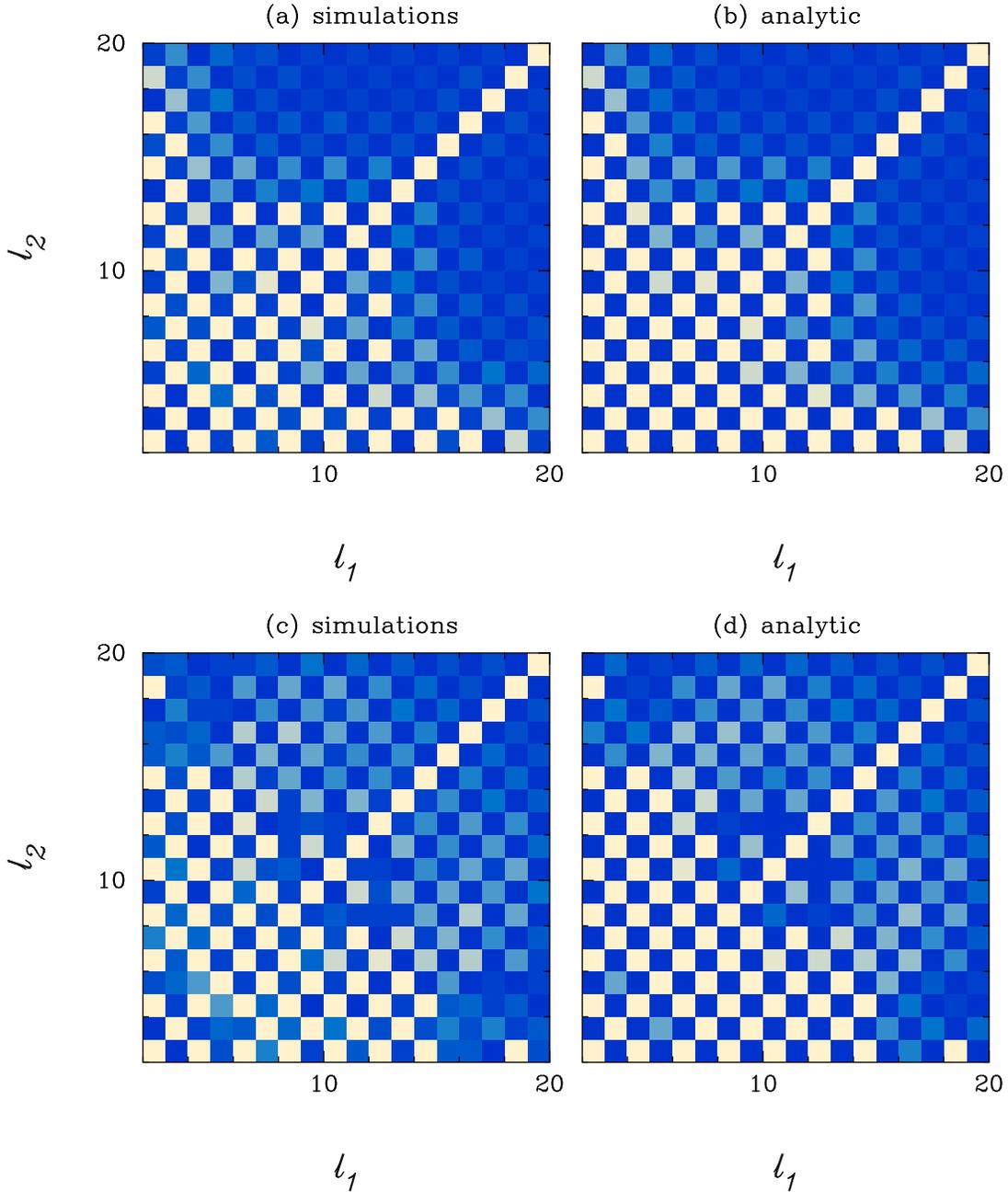


\vskip 6.8 truein

\includegraphics{pgcoup30a.ps}
\includegraphics{pgcoup30b.ps}

\caption
{As Figure 3, but now comparing with the exact 
expression for the covariance matrices  (equation \ref{V6}).}

\label{figure4}

\end{figure*}

 The procedure for estimating an accurate covariance matrix for a PCL estimator is
therefore as follows:

\noindent
(i) Adopt a (smooth) theoretical model for the power spectrum $C_\ell$. (This can
be done iteratively, with an intermediate parameter estimation step if required).

\noindent
(ii) Compute the covariance matrix for PCL estimates from the full resolution
map using the approximate expression (\ref{V2}).

\noindent
(iii) Use the exact expression (\ref{V6}) to compute the covariance matrix at low
multipoles by summation using a lower resolution pixelization.

\noindent
(iv) Replace the elements of the covariance matrices at low multipoles ($\ell_1 \le \ell_{\rm J}$,
$\ell_2 \le \ell_{\rm J}$) computed in step (ii) with the elements computed in step (iii).

There are a number of subtleties involved in this procedure. For
example, what resolution should be used in step (iii)? How
does one choose the `joining' multipole $\ell_{\rm J}$ in step (iv)?
What should one use for the final covariance matrix in the ranges
($\ell_1 \le \ell_{\rm J}$, $\ell_2 > \ell_{\rm J}$) and ($\ell_2 \le
\ell_{\rm J}$, $\ell_1 > \ell_{\rm J}$) which are not computed in step
(iii)? In the example discussed in Section (4), the joining multipole $\ell_J$ is
simply chosen by trial and error so that the covariances matrices computed in
steps (ii) and (iv) agree to high accuracy (a few percent or
better). The components of the covariance matrices at ($\ell_1 \le
\ell_{\rm J}$, $\ell_2 > \ell_{\rm J}$) and ($\ell_2 \le \ell_{\rm
J}$, $\ell_1 > \ell_{\rm J}$) are simply retained, but their values
are so low that for most purposes it would make no difference if these
values were set to zero. None of these subtleties are critical, and
the method is so fast that many choices can be explored to assess the
accuracy of the method. As long as the sky cut is not too large, the method
described here can return a deconvolved PCL estimate $\hat C^p_\ell$,
together with an accurate covariance matrix. If the sky cut is large, 
then the method can return a PCL estimate $\tilde C^p_\ell$
(or `regularised' estimate, equation (\ref{PCL11}))  together
with an accurate covariance matrix. Including uncorrelated instrumental 
noise poses no fundamental problems, and is discussed in Section 5.

\section{Maximum likelihood Estimation}

\subsection{NRML Estimators}

In the NRML method (Bond \etal, 1998) a set of parameters $a_p$,
which may be estimates of the power spectrum $C_\ell$,  is determined
from the likelihood function (\ref{I1}) by Newton-Raphson iteration.
First, guess a set of values for the parameters $a_p$ and then refine
these by adding a correction
\beglet
\begin{equation}
 \delta a_p = 
  \sum_{p^\prime}  \ F_{p p^\prime}^{-1}  {1 \over 2}{\rm Tr} 
\left [ \left (x x^T - C \right ) \left ( C^{-1} {
\partial C \over \partial a_p } C^{-1} \right) \right], \label{NR1}
\end{equation}
where $F$ (the Fisher matrix)  is the expectation value of the curvature matrix 
\begin{equation}
F_{p p^\prime} = - \left \langle
{\partial^2 {\cal L} \over \partial a_p \partial a_{p^\prime}} \right \rangle
 = {1 \over 2 }  {\rm Tr} 
\left [  C^{-1} {\partial C \over \partial a_p} C^{-1} {\partial C \over \partial
a_{p^\prime}} \right ]. \label{NR2}
\end{equation}
\endlet
This leads to a new set of values $a_p$ and the computation can be repeated until
the estimates $a_p$ converge. An estimate of the covariance matrix for the parameters
$a_p$ is given by the
inverse of the Fisher matrix at the final iteration
\begin{equation}
 \langle \delta a_p \delta a_{p^\prime} \rangle =  
F_{p p^\prime}^{-1} . \label{NR3}
\end{equation}
An alternative, and computationally more expensive NRML method uses the full
curvature matrix instead of $F$ in equation (\ref{NR1}) (Borrill 1999b). The NRML
estimator has been used widely in the analysis of CMB experiments, for example,
MAXIMA-1 (Hanany \etal, 2000), BOOMERANG (de Bernardis \etal,  2000), WMAP
(Hinshaw \etal, 2003) and to galaxy clustering (Efstathiou and Moody 2001).

There are, however, some disadvantages in using the NRML method. If
the method is to be used to estimate power spectra from a map with a large
sky cut, the Fisher matrix
$F_{\ell \ell^\prime}$ will be numerically singular ({\it i.e.} will
have a high condition number, see Press \etal, 1992).
 In this case, the iterations (\ref{NR1}) will not
converge. This problem can be overcome by solving for band-power
averages of the $C_\ell$ (Bond \etal, 1998). However, band-power
averaging  involves a loss of information and can produce biases in the
final estimates of the power spectra\footnote{Even if there is no
band-averaging, the final estimates of $C_\ell$ will be biased since
the method iterates to a maximum likelihood solution that is only
assymptotically unbiased (Bond \etal, 1998).}, depending on the choice
of bands. Cosmological parameters are often estimated from the power
spectrum estimates, $C^e_\ell$, by minimising a $\chi^2$,
\begin{equation}
 \chi^2 (q_p) = \sum_{\ell \ell^\prime} (C^T_\ell(q_p) - C^e_\ell) 
 F_{\ell \ell^\prime} (C^T_{\ell^\prime}(q_p) - C^e_{\ell^\prime}), \label{NR4}
\end{equation}
with respect to the parameters $q_p$, where $C^T_\ell$ is the
theoretical power spectrum. If the Fisher matrix (\ref{NR2}) is used
in equation (\ref{NR4}) then the parameters $q_p$ will be biased (Bond
\etals 1998; Oh \etals 1999). This is because a low estimated value of
$C^e_\ell$ will be assigned a low variance. This problem can be
partially alleviated by smoothing the final estimates $C^e_\ell$ and
re-computing the Fisher matrix (Oh \etals 1999). However, it is easy
to think of cases where such a procedure would fail catastrophically,
for example, for the low quadrupole amplitude measured by WMAP
(Bennett \etals 2003). To avoid bias, what is wanted in equation
(\ref{NR4}) is the Fisher matrix appropriate to the estimator for the
theoretical model of interest, {\it i.e.} $F_{\ell
\ell^\prime}(C^T_\ell)$, and preferably an accurate model of the
likelihood function rather than the simple $\chi^2$ of equation
(\ref{NR4}). Clearly, it is not feasible to evaluate equation
(\ref{NR2}) for every set of physical parameters $q_p$.  One solution
is to find a model for $F_{\ell \ell^\prime}$ which can be rescaled
(or recallibrated) to a good approximation for any given theoretical
model (see {\it e.g.} Verde \etals 2003). This approach is, in part,
motivation for the hybrid estimator developed in Sections 4 and 5. For
another approach to this problem, see Gupta and Heavens 2002).

Finally, since the NRML method iterates to a ML solution, the final estimates
of $C^e_\ell$ are not easily expressible in terms of the data vector ${\bf x}$.
In contrast, the QML estimator discussed in the next Section, as its name
implies, is expressible in terms of products $x_i x_j$ of the data vector. This
property is critical in defining the hybrid estimator of Sections 4 and 5.

\subsection{QML Estimators}

QML estimators have been used extensively by Tegmark and
collaborators ({\it e.g.} Tegmark \etals 1998; Hamilton, Tegmark and
Padmanabhan, 2000; Tegmark \etals 2002a; Tegmark, Hamilton and Xu,
2002b). Most other authors have used the NRML methods described in the
previous sub-section, giving the impression that QML estimators are in
some sense inferior to NRML estimators. One of the purposes of this
sub-section is to show that this is impression is unjustified. As in
previous Sections, instrumental noise is ignored to simplify the
discussion. Instrumental noise is discussed in Section 5.2.

If the data vector $x_i$ is composed of Gaussian random variates, 
a minimum variance quadratic estimator of the power spectrum  is
\beglet
\begin{equation}
 y_{\ell} = x_i x_j E^\ell_{ij}, \label{ML1a}
\end{equation}
where 
\begin{equation}
 E^\ell = {1 \over 2}C^{-1} {\partial C \over \partial C_\ell} C^{-1}. \label{ML1b}
\end{equation}
\endlet 
(Tegmark 1997). Evidently, a functional form for $C_\ell$ ($C^{\rm in}_\ell$) must
be assumed to compute the matrix (\ref{ML1b}). However, the
expectation value of $y_\ell$ provides an {\it unbiased} estimator of
the true $C_\ell$ whatever the assumed form for $C^{\rm in}_\ell$,
\begin{equation}
 \langle y_{\ell} \rangle  = F_{\ell \ell^\prime} C_{\ell^\prime}, \label{ML2}
\end{equation}
where $F_{\ell \ell^\prime}$ is the Fisher matrix of equation (\ref{NR2}) with
$a_p$ replaced by $C_\ell$. As it stands, the vector $y_\ell$ will (usually)
look very different to the power spectrum. However, we can recover an estimate
that has a similar shape to the true power spectrum simply by rescaling
\begin{equation}
 \tilde C^q_{\ell}   =  y_\ell / \sum_{\ell^\prime} F_{\ell \ell^\prime}. \label{ML3}
\end{equation}

The covariance matrix for the estimates $y_\ell$ is 
\begin{equation}
\langle y_\ell y_{\ell^\prime} \rangle 
- \langle y_\ell \rangle \langle y_{\ell^\prime} \rangle 
= 2 {\rm Tr} \left [ C E^{\ell} 
C E^{\ell^\prime} \right ],  \label{ML4} 
\end{equation}
where the $C$'s appearing in equation (\ref{ML4}) are the true covariance matrices.
If the assumed form of $C_\ell$ is a good approximation to the
true power spectrum, then equation (\ref{ML4}) simplifies to
\begin{equation}
\langle \Delta y_\ell \Delta y_{\ell^\prime} \rangle = F_{\ell \ell^\prime}. \label{ML5}
\end{equation}
To summarise, if the initial guess is a poor approximation to the true
 $C_\ell$, equation (\ref{ML2}) provides an unbiased estimate of the
 power spectrum provided that the guess for $C_\ell$ is used to
 compute the covariance matrices in the expressions for $E$ and $F$
 (equations (\ref{ML1b}) and (\ref{NR2})).  If the initial guess for
 $C_\ell$ is poor, the resulting estimates $y_\ell$, although
 providing unbiased estimates of $C_\ell$, will not be the ML solution
 and their covariances will be given by equation (\ref{ML4}); the
 errors on the QML estimates, although accurate, will not be as small
 as those of a ML solution.  If the initial guess for $C_\ell$ is
 close to the true answer, then the estimates of $y_\ell$ will be
 close to the ML solution and their covariances will be given accurately  by
 equation (\ref{ML5}).  In practice, finding an acceptable initial
 guess for $C_\ell$ is unlikely to be a problem since the QML
 estimates at low multipoles are almost independent of the initial
 guess (see Section 3.3) and the CMB power spectrum at intermediate
 multipoles is already well estimated from WMAP.  Henceforth we will
 assume that the initial guess for $C_\ell$ is close enough to the
 true answer that equation (\ref{ML5}) applies.

If the Fisher matrix is non-singular (as is the case for a narrow
Galactic cut) then equation (\ref{ML2}) can be inverted to provide an
unbiased QML estimate of $C_\ell$,
\begin{equation}
  \hat C^q_{\ell}  = F^{-1}_{\ell \ell^\prime} y_{\ell^\prime}
, \label{ML6}
\end{equation}
with covariance matrix
\begin{equation}
 \langle \Delta \hat C^q_{\ell} \Delta \hat C^q_{\ell^\prime}
\rangle  = F^{-1}_{\ell \ell^\prime}.  \label{ML7}
\end{equation}
The matrix $F$ is therefore an approximation to the Fisher matrix for
the estimates $\hat C^q_{\ell}$.

With this formulation, there is a close analogy between the PCL
estimates $\tilde C^p_\ell$ and $\hat C^p_\ell$ (equations
(\ref{PCL7}) and (\ref{PCL10})) and the QML estimates $\tilde
C^q_\ell$ and $\hat C^q_\ell$ (eqations (\ref{ML3}) and (\ref{ML6})).
Both of the estimates $\tilde C^p_\ell$ and $\tilde C^q_\ell$ are related
to the true power spectrum by convolutions. If a narrow Galactic cut
is used, then these estimates can be deconvolved to form $\hat C^p_\ell$
and $\hat C^q_\ell$. However, this deconvolution step is not essential, since
$\tilde C^p_\ell$ and $\tilde C^q_\ell$ contain the same information as
the deconvolved estimates $\hat C^p_\ell$ and $\hat C^q_\ell$.

\subsection{Fisher matrix for low multipoles}

Suppose that the data vector $x_i$ consists of the spherical harmonic
coefficients $\tilde a_{\ell m}$ measured on the cut sky.  These
coefficients are related to the true harmonic coefficients by the
coupling matrix $K_{{\ell m} {\ell^\prime m^\prime}}$ (equation
\ref{PCL3}).  As discussed in Section 2.1, on a cut sky some
combinations of the $a_{\ell m}$ define `null modes' that lie within
the cut and hence the matrix $K$ will be singular.  However, if the
Galactic cut is relatively narrow, the coefficients $\tilde a_{\ell m}$
at low multipoles will be weakly correlated with any of the $a_{\ell
m}$ that define `null modes' (Figure 2). In this case, the expansion
(\ref{PCL3}) can be truncated at finite ($\ell, m$) and ($\ell^\prime,
m^\prime$), and the truncated matrix $\tilde K$ for the cut sky can be
inverted. (The invertibility of $\tilde K$ can be used to construct
orthogonal functions on the cut sky from the spherical harmonics
$Y_{\ell m}$, G\'orski 1994, Mortlock \etals 2002).

If $\tilde K$ is invertible, then we can reconstruct the approximations to the
actual $a_{\ell m}$ coefficients
on the uncut sky by performing the matrix inversion $a \approx \tilde K^{-1} \tilde a$. The
QML estimates of $C_\ell$ (in the absence of instrumental noise) then reduce to
\begin{equation}
\hat   C^q_{\ell} = {1 \over (2 \ell + 1)} \sum_m \vert a_{\ell m} \vert^2. \label{ML8}
\end{equation}
and so return almost the {\it exact} value of $C_\ell$ for our
particular realization of the CMB sky in the presence of a Galactic
cut (see also Efstathiou 2004). The Fisher matrix for the QML
estimates is therefore simply
\begin{equation}
  F_{\ell \ell^\prime} =  {(2 \ell + 1) \over 2 C_\ell^2} 
\delta_{\ell \ell^\prime},  \label{ML9}
\end{equation}
reflecting cosmic variance.
In other words, the QML  estimates $\hat C^q_{\ell}$ give
cosmic variance limited estimates of the power spectrum
{\it independent} of the size of a Galactic cut, provided that the cut is not too
large. (For example, equation (\ref{ML9}) is a good approximation at low multipoles
for the Kp2 sky cut defined by the WMAP team but begins to break down for the
Kp0 and larger sky cuts, see Efstathiou 2004). The usual formula 
\begin{equation}
  \langle \Delta C_\ell^2 \rangle = {2 C^2_{\ell} \over (2 \ell + 1) 
f_{\rm sky}}, \label{ML10}
\end{equation}
({\it e.g.} Knox 1995) where $f_{\rm sky}$ is the fraction of the sky
sampled by the experiment, clearly does not apply at low multipoles
and one can think of the $f_{\rm sky}$ factor in (\ref{ML10}) as
reflecting the loss of information at high multipoles from the missing
`null modes' that lie within the Galactic cut.

\begin{figure}

\vskip 3.2 truein

\includegraphics{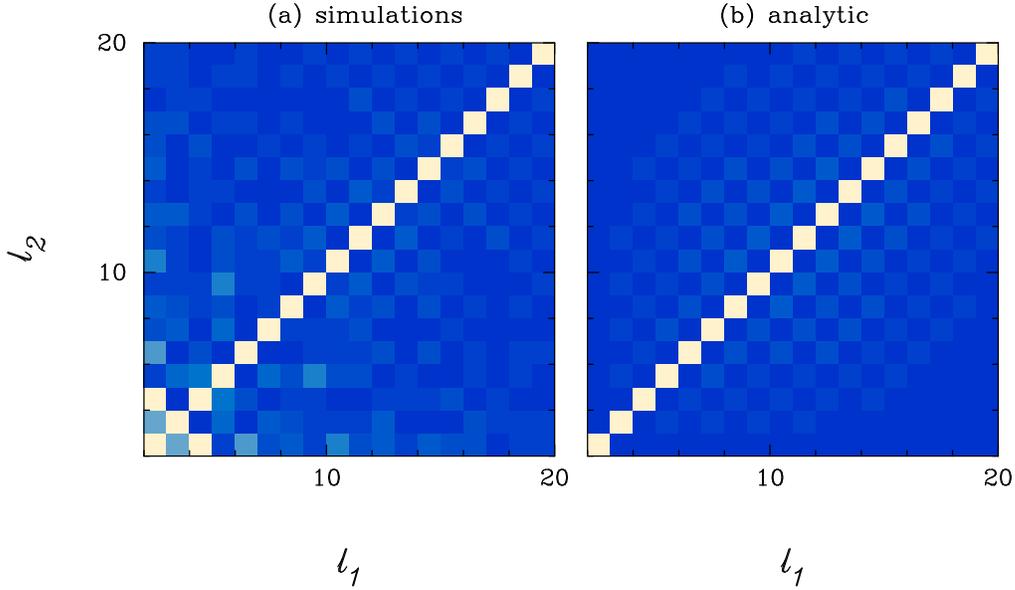}

\caption
{The figure to the left shows the covariance matrix for the 
quadratic estimator $\hat C^q_{\ell}$ (equation \ref{ML6}) computed
from $2 \times 10^5$ simulations. The figure to the right shows
the inverse of the Fisher matrix (equation \ref{ML7}).}

\label{figure5}

\end{figure}

The results of this subsection are borne-out by a set of $2 \times
10^5$ numerical simulations. In these simulations we apply the
quadratic estimator (\ref{ML1a}) to the same set of simulations used
to construct Figures 3 and 4, using the pixel $\Delta T_i$ values as
the data vector $x_i$.  Figure 5 shows the resulting covariance matrix
for the estimates $\hat C^q_\ell$ compared to the inverse of the
Fisher matrix. There is some low-amplitude off-diagonal structure in
both of the plots (as a consequence of the small coupling with `null
modes') but this is small and negligible for most purposes. The
results from the simulations agree with equation (\ref{ML9}) to within
a few percent at multipoles $\ell \simlt 5$) and rise smoothly to
match approximately with equation (\ref {ML10})  by $\ell =
30$ (see also Figure \ref{figure14}a below).  Note that for the
simulations used here, $f_{\rm sky} = 0.83$. The discrepancy between
the analytic and simulation results at low $\ell$ in Figure 5 is
simply a consequence of the finite number of simulations. The
off-diagonal elements of the covariance matrix estimated from the
simulations at low multipoles are small compared to the diagonal
elements (note that the grey-scale in Figures 2-5 was chosen to
accentuate small amplitude structure) and become smaller as the number
of simulations is increased.

\begin{figure}
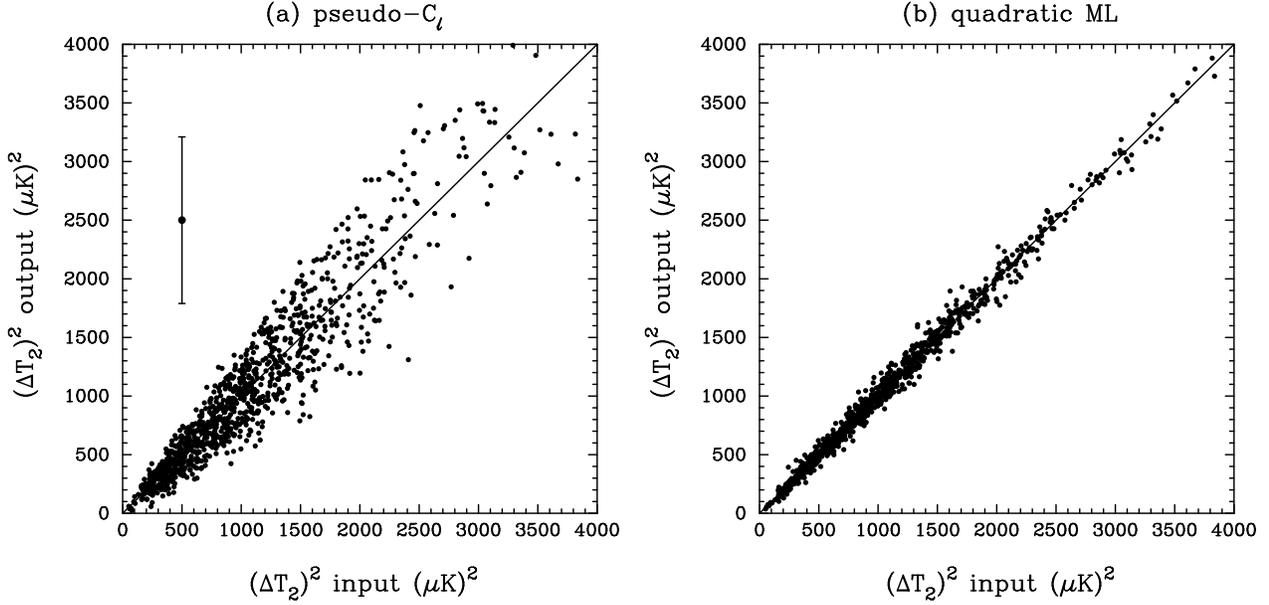


\vskip 3.1 truein

\includegraphics{pgcl2a.ps}
\includegraphics{pgcl2b.ps}

\caption
{The figure to the left  shows the quadrupole amplitude estimated
from the PCL estimator (ordinate)  plotted against the quadrupole amplitude 
of the simulation used to generate the map (abscissa). (Points for 1000 simulations
are shown). The figure to the right  shows the equivalent plot for the QML
estimator. The error bar in Figure (\ref{figure6}a) shows the cosmic variance
error for the mean value of $(\Delta T_2)^2 = 1123 \; (\mu {\rm K})^2$ for the
input simulations.}

\label{figure6}

\end{figure}

 The behaviour of the QML estimator differs from the PCL
estimates $\hat C^p_{\ell}$, which are highly correlated at low
multipoles ({\it e.g.} Figure 4) and which have a variance that
significantly exceeds cosmic variance. As an example, Figure 6 compares
the amplitudes of the quadrupole estimates, expressed as
\begin{equation}
   (\Delta T_\ell)^2 =  { 1 \over 2 \pi}  \ell (\ell + 1) C_{\ell}, 
\label{ML11}
\end{equation}
compared to the input quadrupole amplitude in each simulation. The
expectation value of the quadrupole amplitude for the simulations is
$(\Delta T_2)^2 = 1123 \; (\mu {\rm K})^2$ (after smoothing with the
Gaussian beam function) and so the cosmic variance is expected to lead
to a dispersion of $710 \; (\mu {\rm K})^2$. Figure 6, shows that the
`estimator induced' dispersion in the PCL case ($262 \; (\mu {\rm
K})^2$) is a significant fraction of the cosmic variance . For the QML
case, the `estimator induced' dispersion is much smaller ($68 \; (\mu
{\rm K})^2$). These numbers depend on the size of the Galactic cut,
but it is clear that for realistic Galactic cuts the QML estimator
returns amplitudes for low multipoles that are close to the true
values for our particular realisation in the sky. This can be
important, {\it e.g.} the WMAP quadrupole estimated using a PCL
estimator has an amplitude of only $(\Delta T_2)^2 = 123 \; (\mu {\rm
K})^2$, much smaller than the amplitude expected in the concordance
$\Lambda$CDM cosmology (Bennett \etals 2003; Spergel \etals 2003). In contrast,
the QML estimator applied to the WMAP internal linear combination map gives
a quadrupole amplitude closer to $200 \; (\mu {\rm
K})^2$ (Efstathiou 2004).

\subsection{Fisher matrix at high multipoles}

In the pixel domain, the signal covariance matrix is
\begin{equation}
S_{ij} = \sum_\ell {(2 \ell + 1) \over 4 \pi} w_i w_j
C_\ell P_{\ell} (\theta_{ij}) = \sum_{\ell m} C_\ell w_i w_j Y_{\ell m}
({\bf \theta}_i) Y^*_{\ell m} ({\bf \theta}_j), \label{HF1}
\end{equation}
where  $w_i$ is the (arbitrary) pixel weight function introduced
in equation (\ref{PCL1}). If the data cover the whole sky (and the weight function
$w_i$ is everywhere non-zero), then from the orthogonality of the spherical
harmonics, the inverse signal covariance matrix is given by
\begin{equation}
S_{ij}^{-1} = \sum_{\ell m} {\Omega_i \Omega_j \over C_\ell w_i w_j} 
Y^*_{\ell m}({\bf \theta}_i) Y_{\ell m} ({\bf \theta}_j). \label{HF2}
\end{equation}
Although equation (\ref{HF2}) applies strictly only for the complete
sky, it will be a good approximation for a cut sky with $i$ and $j$
restricted to the pixels outside the cut, provided that the total number of
pixels in the map is very much greater than the number of pixels close
to the boundary. (This condition is satisfied in all of the examples
discussed in this paper). The product $C^{-1}\partial C/ \partial
C_\ell$ in (\ref{ML1b}) is then
\begin{eqnarray}
S_{ij}^{-1} {\partial S_{jk} \over \partial C_\ell} &=& \sum_{j} \sum_{m
 \ell_1 m_1} { \Omega_j \Omega_j \over C_{\ell_1} w_i w_j} Y_{\ell_1
 m_1} (\theta_i) Y^*_{\ell_1 m_1} (\theta_j) \; w_j w_k Y_{\ell m}
 (\theta_j) Y^*_{\ell m} (\theta_k) \nonumber \\
 & \approx & {1 \over C_\ell} \sum_m \Omega_i {w_k \over w_i} Y_{\ell m} (\theta_i) Y^*_{\ell m} (\theta_k). \label{HF3}
\end{eqnarray}
The Fisher matrix (\ref{NR2}) is therefore
\begin{equation}
F_{\ell \ell^\prime} \approx {1 \over 2 C_\ell C_{\ell^\prime}} \sum_{ik} \sum_{m m^\prime}
\Omega_i \Omega_k Y_{\ell m} (\theta_i) Y^*_{\ell m} (\theta_k) Y_{\ell^\prime m^\prime}
(\theta_k) Y^*_{\ell^\prime m^\prime} (\theta_i). \label{HF4}
\end{equation}
Notice that weight factors $w_i$ have cancelled and do not appear in
equation (\ref{HF4}). This is what is expected; if the data
vector $x_i$ is multiplied by arbitrary weights and a maximum
likelihood estimator applied to estimate $C_\ell$, then neither the
estimates $\hat C^q_\ell$ nor their covariance matrix should depend on
the weights. The derivation leading to equation (\ref{HF4}) differs
from that in Appendix D.1.2 of Hinshaw \etals (2003) which assumes
equal weight per pixel. The final answer does, however, agree with equation
(D19) of Hinshaw \etals  (2003). 

Evaluating the summations in equation (\ref{HF4}), we find
\begin{equation}
  F_{\ell \ell^\prime} \approx {(2 \ell + 1) (2 \ell^\prime + 1) \over 2  C_{\ell} C_{\ell^\prime}}  G_{\ell \ell^\prime} = {(2 \ell + 1) \over 2  C_{\ell} C_{\ell^\prime}}  M_{\ell \ell^\prime},   \label{HF5}
\end{equation}
where the matrices $G_{\ell \ell^\prime}$ and $M_{\ell \ell^\prime}$ are the coupling matrices defined by equations (\ref{PCL8}) and (\ref{PCL9}) {\it evaluated for unit weight per
pixel} over the uncut sky. 

  Notice that for high multipoles, equation (\ref{V4}) for the covariance matrix of
the deconvolved PCL estimates is equal to
\begin{equation}
 \langle \Delta \hat C^p_\ell  \Delta \hat C^p_{\ell^\prime} \rangle  \approx {2 C_\ell
C_{\ell^\prime} \over (2 \ell + 1)} M^{-1}_{\ell \ell^\prime},
\label{HF6}
\end{equation}
which is equal to the inverse of equation (\ref{HF5}). It therefore
follows that: (i) in the signal dominated limit and (ii) for high
multipoles ($ \ell$ greater than the width of the window function
$\tilde W_\ell$), the deconvolved PCL estimate {\it with equal weight
per pixel is optimal and statistically equivalent to a maximum-likelihood
solution}. In the signal-dominated limit, there is therefore nothing
to be gained by solving the ${\cal O}(N_d^3)$ computational problem to
apply a ML method, since the much faster PCL estimate will give a
result that is statistically equivalent.

\section{Hybrid estimator}

\begin{figure*}

\vskip 6.0 truein

\includegraphics{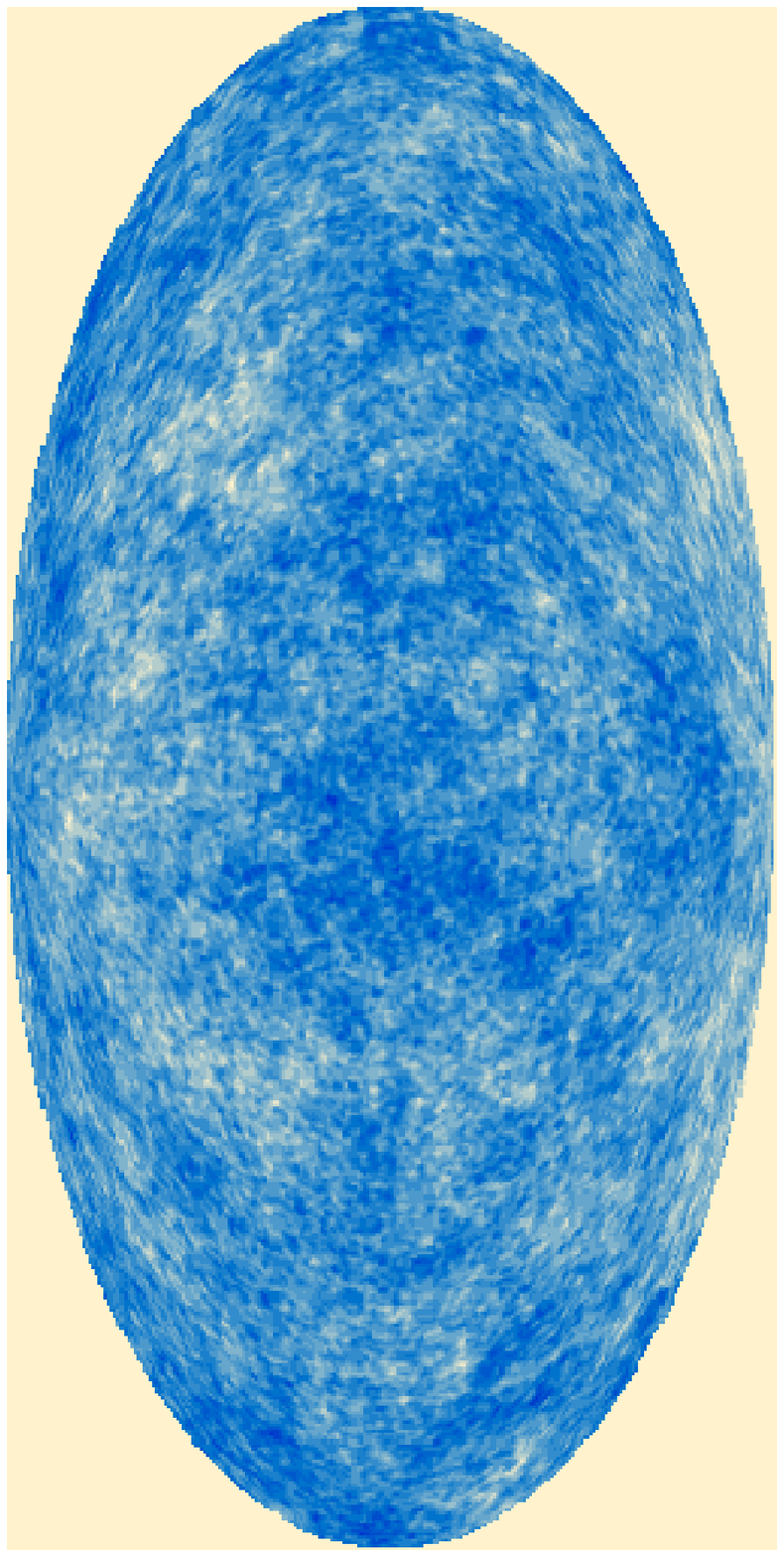}
\includegraphics{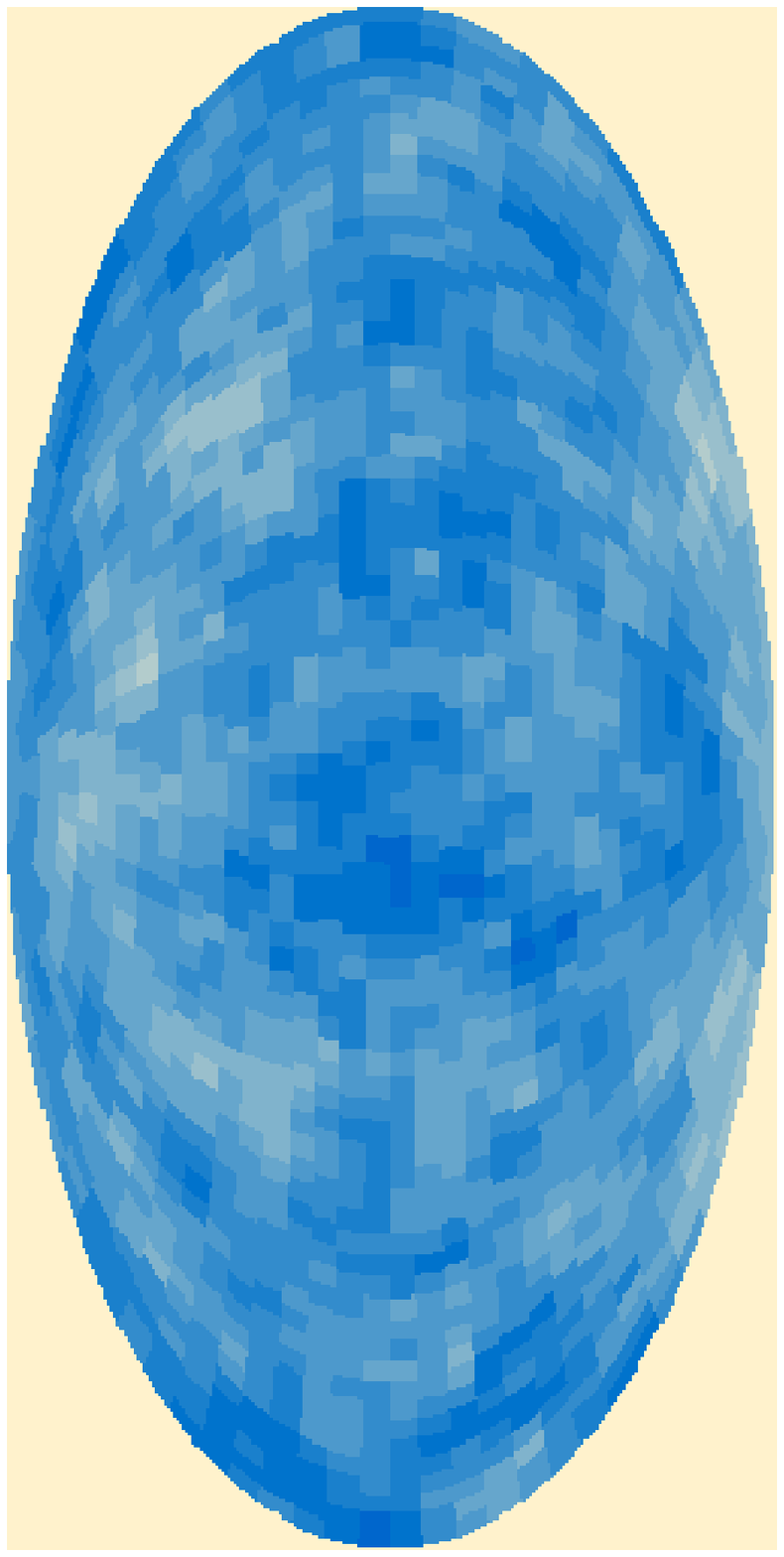}


\caption
{The upper figure shows a simulated CMB sky with smoothing
$\theta_s=1^\circ$ pixelized into $0.5^\circ \times 0.5^\circ$
pixels using an `igloo' type
pixelization scheme ($164828$ pixels in total).  The lower figure
shows the same map smoothed to $\theta_s = 5^\circ$ and pixelized
into $1632$ pixels of area $5^\circ \times 5^\circ$.}

\label{figure7}

\end{figure*}

Having established that the PCL estimator is very close to optimal at
high multipoles, it is clear that a fast and almost optimal power
spectrum estimate can be derived by combining a PCL estimate with a
QML estimate applied to a degraded low resolution map. As an example,
consider the maps shown in Figure \ref{figure7}. The map in the upper
panel is one of the simulations used for the tests described in
Section 2.1 (pixel size\footnote{The pixel areas are identical on
lines of constant latitude, but vary slightly with latitude.}
$\theta_c = 0.5^\circ$, $\theta_s = 1^\circ$).  The map in the lower
panel was constructed from the same $a_{\ell m}$ coefficients with
pixel size $\theta_c = 5^\circ$ and smoothed to
$\theta_s=5^\circ$. Figure \ref{figure7} plots the maps over the whole
sky, but as in previous Sections a cut of $\pm 10^\circ$ above and
below the notional Galactic plane has been applied in the tests
described below.  There are $1385$ active pixels outside Galactic cut
in the low resolution map and so it takes only a few seconds to
compute all of the ${\cal O}(N_d^3)$ operations required for the QML
estimator.

From these maps, two estimates of the power spectrum can be formed,
$\hat C^p_\ell$ (up to a maximum multipole $\ell = \ell_p$) and $\hat C^q_\ell$ 
(up to maximum multipole $\ell = \ell_q$), together with  accurate estimates of
their covariance matrices,  as discussed in previous Sections.
The cross-correlation between the QML and PCL estimates at
low multipoles is given by 
\begin{equation}
\langle \Delta \tilde C^p_\ell \Delta  \hat C^q_{\ell^\prime} \rangle \approx 
{2 C_{\ell^\prime}^2
\over (2 \ell^\prime + 1)} M_{\ell \ell^\prime}, \qquad
\langle  \Delta \hat C^p_\ell  \Delta \hat C^q_{\ell^\prime} \rangle \approx {2 C_{\ell^\prime}^2
\over (2 \ell^\prime + 1)} \delta_{\ell \ell^\prime},  \label{H1}
\end{equation}
where we have assumed the approximate expression for the QML estimator of
equation (\ref{ML8}). Equation (\ref{H1}) has been used in the analysis
described below and is a perfectly adequate approximation for $\ell \simlt 30$.
If necessary, a more accurate expression for the cross-correlations can be
evaluated from a sum over pixels, as in the analysis of the PCL covariance
matrix described in  Section 2.2:
\beglet
\begin{equation}
\langle  \Delta \tilde C^p_\ell \Delta  y_{\ell^\prime} \rangle = 
{2 \over (2 \ell + 1)} \sum_{mpq} z_{plm} z^*_{qlm} E^{\ell^\prime}_{pq},  \label{H2a}
\end{equation}
where
\begin{equation}
 z_{jlm} = \sum_i C(\theta_{ij}) w_i \Omega_i Y_{\ell m}(\theta_i).  \label{H2b}
\end{equation}
\endlet

\begin{figure*}

\vskip 4.0 truein

\includegraphics{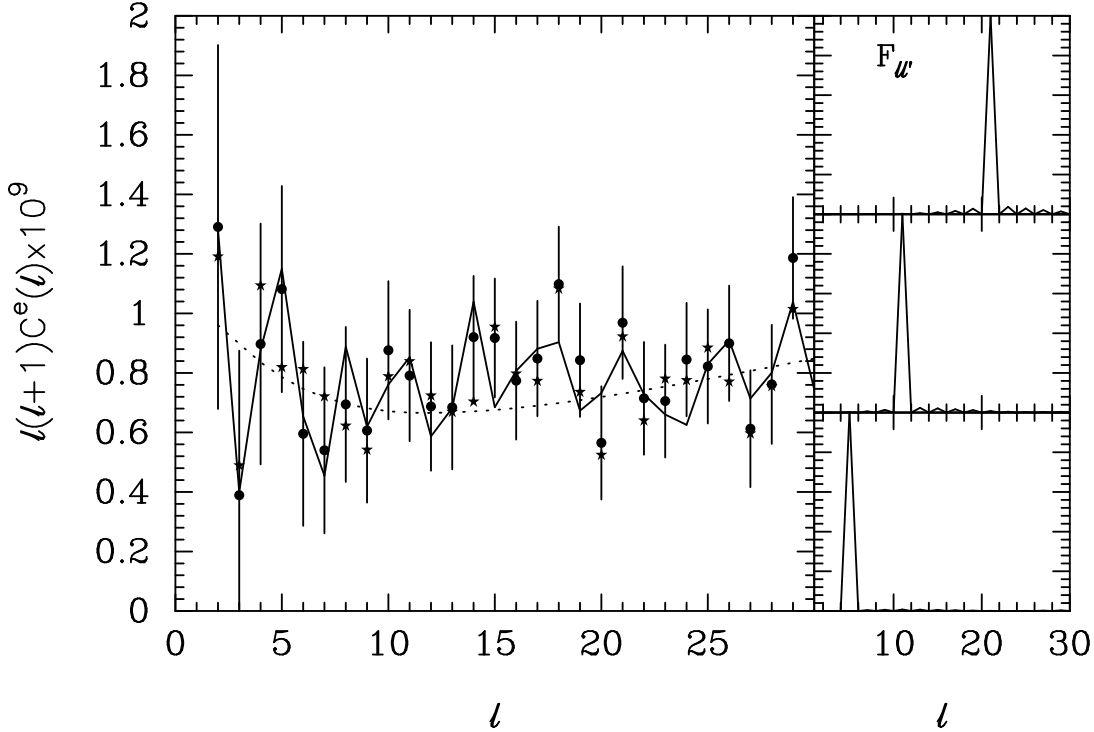}

\caption
{The filled circles show the results of applying the QML estimator
to the low resolution map shown in Figure \ref{figure7}. The error bars
are computed from the diagonal components of the inverse of the Fisher
matrix (equation \ref{NR2}). Three rows of the Fisher matrix are
shown in the panels to the right. The stars show the deconvolved
PCL estimates $\hat C^p_\ell$ determined from the high resolution
map shown in Figure \ref{figure7}. The dotted line shows the fiducial $\Lambda$CDM
power spectrum and the solid line shows the actual
power spectrum computed from the $a_{\ell m}$ coefficients used to generate
the maps. A sky cut of $\pm 10^\circ$ has been used for the estimated power
spectra and the QML estimates have been corrected to a beam smoothing of
$1^\circ$ as used in the high resolution maps.}

\label{figure8}

\end{figure*}

Given expressions for the covariances and the cross-covariances, we can combine the 
two estimates $\hat C^q_\ell$ and $\hat C^p_\ell$ 
into a single data vector $\hat C_{\alpha \ell}$ ($\alpha \equiv q, p)$
and form the $\chi^2$
\begin{equation}
\chi^2 =    (\hat C_{\alpha\ell_1} - 
\hat C^h_{\ell_1}){\cal F}_{\alpha \ell_1 \beta \ell_2}  (\hat C_{\beta \ell_2} - 
\hat C^h_{\ell_2}),    \label{H3}
\end{equation}
where ${\cal F}_{\alpha \ell_1 \beta \ell_2}$ is the inverse of
the covariance matrix $\langle \Delta \hat C_{\alpha \ell_1} 
\Delta \hat C_{\beta \ell_2} \rangle$.
Minimising equation (\ref{H3})  gives the solution
for the hybrid estimate  $\hat C^h_{\ell}$  
\begin{equation}
  \sum_{\alpha \beta \ell_1} {\cal F}_{\alpha \ell_1 \beta \ell} \hat
  C^h_{\ell_1} =
\sum_{\alpha \beta \ell_1} {\cal F}_{\alpha \ell_1 \beta \ell} \hat C_{\alpha \ell_1},    \label{H4}
\end{equation}
with  covariance matrix
\begin{equation}
\langle  \Delta \hat C^h_{\ell_1} \Delta \hat C^h_{\ell_2} \rangle = 
\left (\sum_{\alpha \beta} {\cal F}_{\alpha \ell_1 \beta \ell_2} \right )^{-1}.
\label{H5}
\end{equation}
The hybrid estimator $\hat C^h_{\ell}$ therefore makes a smooth
transition between the QML estimates at low multipoles and the PCL
estimates at high multipoles (and would be exactly equal to the
quadratic estimates at low multipoles if the Fisher matrix were
precisely diagonal). From the results presented in Sections 2.2 and
3.2, the covariance matrix for the hybrid estimates (\ref{H5}) is
expected to be predominently band-diagonal at all multipoles if the
sky cut is relatively narrow.  Furthermore, since each of the
estimates $\hat C^q_\ell$ and $\hat C^p_\ell$ is an unbiased estimate
of $C_\ell$, it follows from equation (\ref{H4}) that the hybrid
estimator $\hat C^h_\ell$ is unbiased.

\begin{figure*}

\vskip 4.00 truein

\includegraphics{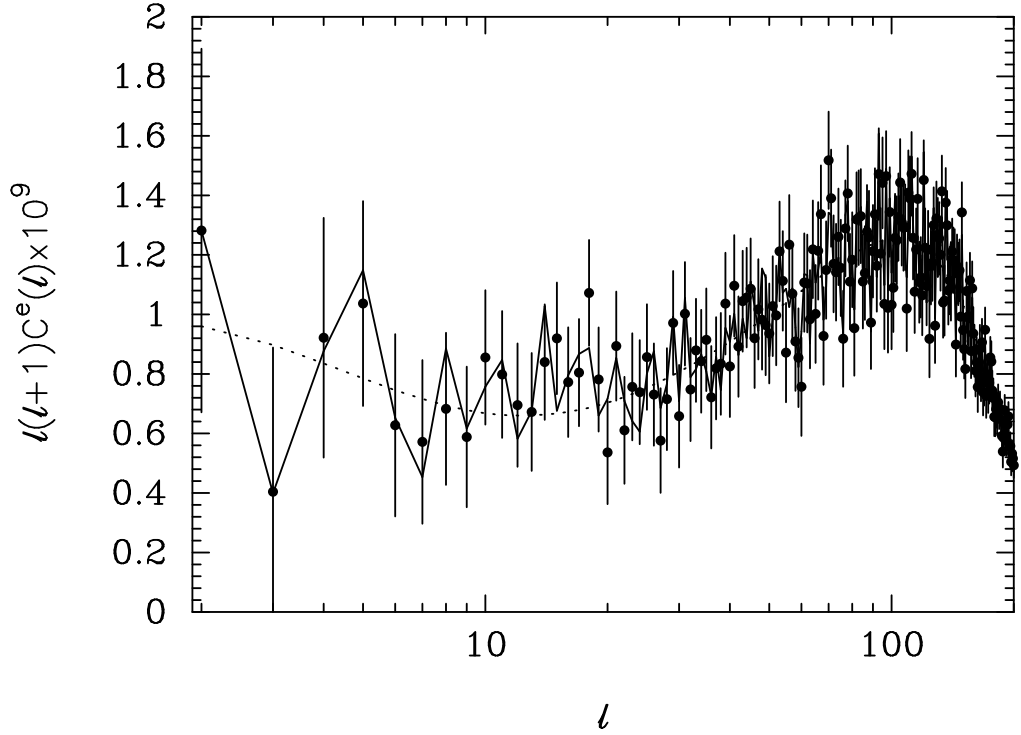}

\caption
{The filled circles show the hybrid estimates $\hat C^h_\ell$ for the maps
shown in Figure \ref{figure7} computed from equation (\ref{H4}). The
error bars were computed from diagonal components of the covariance
matrix (\ref{H5}). As in Figure \ref{figure8}, the dotted line shows
the fiducial $\Lambda$CDM power spectrum and the solid line shows the
actual power spectrum computed from the $a_{\ell m}$ coefficients used
to generate the maps.}

\label{figure9}

\end{figure*}

\begin{figure*}

\vskip 3.1 truein

\includegraphics{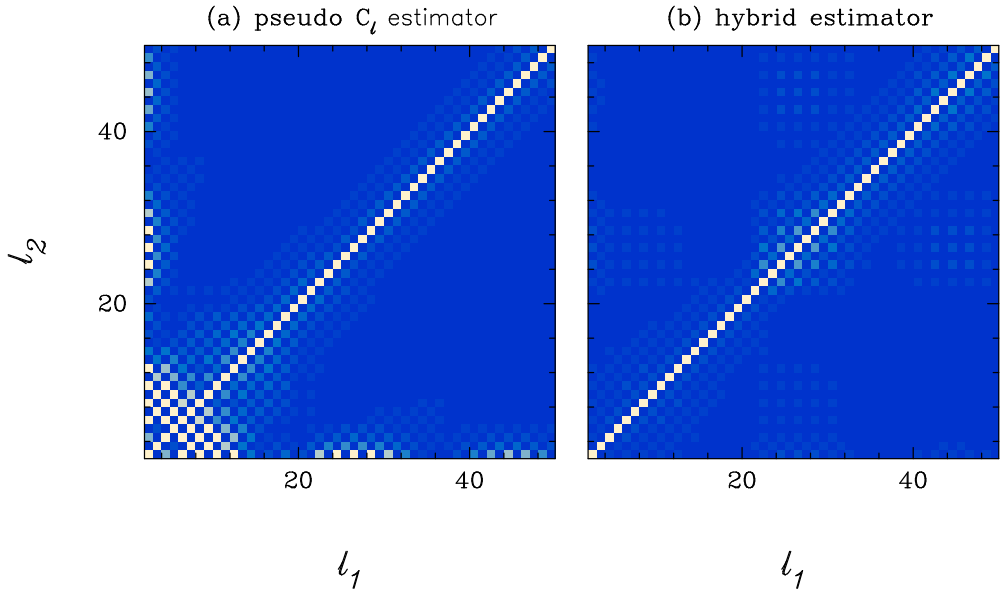}

\caption
{The left hand panel shows the analytic covariance matrix at $\ell \le 50$
for the PCL estimates. The right hand panel shows the
covariance matrix for the hybrid estimates $\hat C^h_{\ell}$
computed from equation (\ref{H5}).}

\label{figure10}

\end{figure*}

To illustrate the hybrid method, Figure \ref{figure8} shows a comparison of
the QML and PCL estimates at low multipoles derived from the maps shown in
Figure \ref{figure7}. Three rows of the Fisher matrix for the QML estimates
are shown in the panels to the right. As expected, for a sky cut of $\pm 10^\circ$,
the QML Fisher matrix is almost exactly diagonal at low multipoles and so the
estimates $\hat C^q_\ell$ agree almost perfectly with the power spectrum
computed from the $a_{\ell m}$ for this particular simulation. By multipoles
$\ell \approx 10$ the QML and PCL estimates become similar.

The hybrid estimate computed from equation (\ref{H4}) are shown in Figure 
\ref{figure9}. In this example, the QML estimates were retained up to $\ell_q = 20$.
For the PCL estimator, the exact expression for the covariance matrix (equation
\ref{V6}) was used for $\ell \le 20$ and the approximate expression (equation \ref{V2}) 
was used for higher multipoles. Neither of these choices are critical, as can be
seen from the close agreement between QML and PCL estimates plotted in Figure
\ref{figure8}. 

The analytic covariance matrices for the PCL estimator ({\it cf}
Figure 2d) and the hybrid estimator (equation \ref{H5}) are plotted in
Figure \ref{figure10} for multipoles $\ell \le 50$. As expected, the
covariance matrix for the hybrid estimator is almost band-diagonal. In
Section 3.4 it was proved that the PCL estimator with equal weight per
pixel is statistically equivalent to the QML estimator if instrumental
noise can be neglected. The hybrid estimates plotted in Figure
\ref{figure9} should therefore be almost indistinguishable from a
brute force ${\cal O}(N_d^3)$ ML analysis of the high resolution map
shown in Figure \ref{figure7}. However, in contrast to the
${\cal O}(N_d^3)$ methods,  computation of the hybrid estimator
and its covariance matrix takes of order seconds on a laptop
computer.

The analysis of the hybrid estimator, as described above, applies if
the estimators $\hat C^p_\ell$ and $\hat C^q_\ell$ can be constructed
from the estimates $\tilde C^p_\ell$ and $\tilde C^q_\ell$, {\it i.e.}
if the Galactic cut is narrow enough. For a large Galactic cut, it may
not be possible to contruct $\hat C^p_\ell$ and $\hat C^q_\ell$. In
this case, one can simply use the hybrid $\chi^2$ of equation
(\ref{H3}) to estimate cosmological parameters (or a suitably modified
function to account for the deviations from Gaussianity at low
multipoles) in place of the $\chi^2$ of equation (\ref{NR4}). Or,
equivalently, one can solve for a suitably regularized hybrid power
spectrum, $R \hat C^h_\ell$, (equation \ref{PCL11}) and its covariance
matrix. There is no requirement to construct a deconvolved hybrid
estimate $\hat C^h_\ell$ and no loss of information in working with a
regularised estimate.

\section{Including  noise}

  Including uncorrelated noise poses no fundamental problems. In the
first two subsections some of the formulae of the previous Sections
are generalised, without detailed mathematical proof, to include
uncorrelated noise. The motivation for applying a hybrid estimator to
noisy data is given in Section 5.3, and an application to a high
resolution simulation with a Planck-type scanning  pattern is described in
Section 5.4. For simplicity, we consider only the case of white noise in this
Section, neglecting the $1/f$-type noise that is expected in a real experiment.
Possible ways of handling $1/f$ noise are discussed briefly in Section 6.

\subsection{PCL estimator including noise}

If noise is included, the PCL estimator (\ref{PCL6}) has expectation value
\begin{equation}
\langle  \tilde C^p_{\ell } \rangle = \sum_{\ell^\prime } C_{\ell^\prime} M_{\ell \ell^\prime} + \langle \tilde N_\ell \rangle, \label{PCLN1}
\end{equation}
({\it e.g.} Hivon \etals 2002) where the noise power spectrum is, in general,
\begin{equation}
\tilde N_\ell  = {1 \over 4 \pi} \sum_{ij} N_{ij} w_i w_j 
\Omega_i \Omega_j P_\ell (\theta_{ij}). \label{PCLN2}
\end{equation}
If the noise matrix is diagonal, $N_{ij} = \sigma_i^2 \delta_{ij}$, equation
(\ref{PCLN2}) simplifies to
\begin{equation}
\tilde N_\ell  = {1 \over 4 \pi} \sum_i \sigma_i^2 w^2_i 
\Omega^2_i . \label{PCLN3}
\end{equation}
To form an unbiased PCL estimator of the power spectrum, then we simply subtract
the noise term, $\tilde C^{\prime p}_\ell = \tilde C^p_\ell -  \tilde N_\ell $,
and compute the deconvolved estimates $\hat C^p_\ell = M^{-1}_{\ell \ell^\prime} \tilde C^{\prime p}_{\ell^\prime}$.

The covariance matrix of $\tilde C^{\prime p}_\ell$ is
\begin{equation}
\langle  \Delta \tilde C^{\prime p}_\ell  \Delta \tilde C^{\prime p}_{\ell^\prime} 
\rangle =  \tilde V^\prime_{\ell \ell^\prime} = \tilde V_{\ell \ell^\prime} + \langle \Delta \tilde C^N_\ell \Delta \tilde C^N_{\ell^\prime} \rangle + \tilde U_{\ell \ell^\prime}, \label{PCLN4}
\end{equation}
where $\tilde V_{\ell \ell^\prime}$ is given by equation (\ref{V2}). The noise
covariance in equation (\ref{PCLN4}) is
\beglet
\begin{equation}  
 \langle \Delta \tilde C^N_\ell \Delta \tilde C^N_{\ell^\prime} \rangle = 
  2 \Xi( \ell, \ell^\prime, \tilde W^N), \label{PCLN5}
\end{equation} 
where 
\begin{equation}
 \tilde W^N_\ell = {1 \over (2 \ell + 1)} \sum_m \vert \tilde
w^{N}_{\ell m} \vert^2, \qquad w^{N}_{\ell m} = \sum_i 
 \sigma_i^2 w_i^2 \Omega_i^2  Y_{\ell m} (\theta_i), \label{PCLN6}
\end{equation} 
\endlet
and the cross-covariance $\tilde U_{\ell \ell^\prime}$ is given by
\beglet
\begin{equation}  
\tilde U_{\ell \ell^\prime} = 4 (C_\ell C_{\ell^\prime})^{1/2} \Xi( \ell, \ell^\prime, \tilde W^{SN}), 
\label{PCLN6a}
\end{equation} 
where 
\begin{equation}
 \tilde W^{SN}_\ell = {1 \over (2 \ell + 1)} \sum_m {\it Re}(
w^{(2)}_{\ell m}  w^{*N}_{\ell m}). \label{PCLN6b}
\end{equation} 
\endlet
The covariance matrix of the deconvolved estimates including noise is
\begin{equation}
 \langle \Delta \hat C^p_\ell  \Delta \hat C^p_{\ell^\prime} \rangle = M^{-1} \tilde V^\prime (M^{-1})^{T},
\label{PCLN7}
\end{equation} 
with $M$ as given in equation (\ref{PCL8}). All of the results quoted in this Section have been
derived under the assumption that the window functions are narrow compared to variations in
$C_\ell$, and the factor $(C_\ell C_{\ell^\prime})^{1/2}$ in equation (\ref{PCLN6a}) has been
written in this way to preserve the symmetry of the matrix $\tilde V^\prime$. An equivalent
result to equation (\ref{PCLN4}) has been derived independently by Chon \etals (2003).

\subsection{QML estimator including noise}

For a general noise matrix $N_{ij}$, we can define an unbiased QML estimate
by redefining $y_\ell$ equation (\ref{ML1a}) as
\begin{equation}
  y^\prime_{\ell} = x_i x_j E^\ell_{ij} - N_{ij}E^\ell_{ij}, \label{QN1}
\end{equation}
(Tegmark 1997). The expectation value of $y^\prime_\ell$ is then given by
equation (\ref{ML2}) with the Fisher matrix as defined in equation
(\ref{NR2}) where $C = S + N$. The covariance matrix of the estimates
$y^\prime_\ell$ are given by equation (\ref{ML4}).

If the noise matrix is diagonal and dominates over the signal, 
$C_{ij} \approx N_{ij} = \sigma_i^2 \delta_{ij}$,
then
\begin{equation}
  E^\ell_{ij} \approx {(2 \ell + 1) \over 8 \pi} {1 \over \sigma_i^2 \sigma_j^2} P_\ell (\theta_{ij}), \label{QN2}
\end{equation}
and the QML estimator is
\begin{equation}
  2y_{\ell} = 2x_i x_j E^\ell_{ij} = \sum_{mij} {x_i \over \sigma_i^2}
{x_j \over \sigma_j^2} Y_{\ell m}(\theta_i) Y^*_{\ell m} (\theta_j) =
(2 \ell + 1) \tilde C^p_\ell,\label{QN3}
\end{equation}
where $\tilde C^p_\ell$ is the PCL estimate using a weight function
$w_i = 1/(\sigma^2_i \Omega_i)$. In other words, if the noise
dominates and is diagonal, the QML estimator is mathematically equivalent to a PCL
estimate computed with inverse variance weighting. In this limit, the
expectation value of the QML estimates $y^\prime_\ell$ is zero and
their covariance matrix is
\beglet
\begin{equation}
 \langle \Delta y^\prime_{\ell_1} \Delta y^\prime_{\ell_2} \rangle
=  F_{\ell_1 \ell_2} = 
  {(2 \ell_1 + 1) (2 \ell_2 + 1) \over 2} 
\Xi(\ell_1, \ell_2, \tilde W^\sigma), \label{QN4}
\end{equation} 
where 
\begin{equation}
 \tilde W^\sigma_\ell = {1 \over (2 \ell + 1)} \vert \tilde
w^\sigma_{\ell m} \vert^2, \qquad w^\sigma_{\ell m} = \sum_i {
1 \over \sigma_i^2} Y_{\ell m} (\theta_i). \label{QN5}
\end{equation} 
\endlet
Equation (\ref{QN4}) agrees with equation (D12) of Hinshaw \etals
(2003).  The covariance matrix of the QML estimates $\hat C^q_\ell =
F^{-1}_{\ell \ell^\prime} y^\prime_{\ell^\prime}$ is given by $F^{-1}$
as in equation (\ref{ML7}) and is identical to the covariance matrix of
equation (\ref{PCLN7}) if the weight function is set to 
$w_i = 1/(\sigma^2_i \Omega_i)$. If the noise is the same in
each pixel and each pixel has the same area ( $\sigma_i^2 = \sigma^2_p$, $\Omega_i = \Omega_p$),  then the diagonal 
components of equation (\ref{QN4}) give the usual approximate formula ({\it e.g.} Knox
1995)
\begin{equation}
   \langle (\Delta \hat C^q_\ell)^2 \rangle \approx  {2 \over (2 \ell +1)}
{ (\sigma^2_p \Omega_p)^2 \over f_{\rm sky}} .  \label{QN6}
\end{equation}

\subsection{Motivation for a hybrid estimator}

The results presented above and in Sections 2 and 3 show that:

\smallskip

\noindent
(I) in the signal dominated limit a PCL estimate
with equal weight per pixel, $w_i=1$,  is statistically equivalent to a ML estimate.

\smallskip

\noindent
(II) in the noise dominated limit, and assuming that the noise matrix
is diagonal, a PCL estimate with inverse variance weighting, $w_i=
1/(\sigma_i^2 \Omega_i)$, is mathematically equivalent to a ML
estimate.

\smallskip

For a realistic experiment such as WMAP, regime (I) is a good
approximation at low multipoles and regime (II) is a good
approximation at high multipoles, but in the intermediate regime it is
difficult to derive a simple analytic formula for the optimal PCL
weights.  To do so, we would need an approximation to the inverse
$C^{-1} = (S + N)^{-1}$, which appears in the Fisher matrix (equation
\ref{NR2}). In the harmonic domain $S$ is sparse while $N$ is not, but
in the pixel domain $N$ is usually sparse while $S$ in not. Hence when
$S$ and $N$ are comparable, the matrix $C$ is not sparse in either
domain and so it is difficult to derive analytic approximations to
$C^{-1}$.  The non-sparseness of $C$ is the fundamental
reason why it is difficult to solve the ${\cal O} (N_d^3)$ problem in
the numerical computation of ML estimators.

One solution, adopted by the WMAP team (Hinshaw \etals 2003), is to 
use a heuristic weighting in the intermediate regime. In the analysis
of the WMAP data, Hinshaw \etals (2003) adopt equal weight per pixel
at $\ell < 200$, inverse variance weighting, $w_i = N_{\rm obs}(i)$
(where $N_{\rm obs}$ is the number of observations, or hit count, of pixel $i$)
at $\ell > 450$, and a heuristic weighting of 
\begin{equation}
     w_i = { 1 \over 1/\langle N_{\rm obs} \rangle + 1/N_{\rm obs} (i)},
\label{HN1}
\end{equation}
in the intermediate range $200 < \ell < 450$, where $\langle N_{\rm
obs} \rangle$ is the average hit count over the unmasked sky. This is
a reasonable solution, but has the disadvantage that the final power
spectrum and associated covariance matrix are contructed from three
estimators with discontinuous joins at $\ell = 200$ and $\ell
=450$. It is worth mentioning here that if the noise matrix is assumed
to be diagonal, the only benefit of solving the full ${\cal O}(N_d^3)$
ML problem at high multipoles is to derive a continuous power
spectrum estimate that is optimal in the intermediate regime.

The hybrid estimator discussed in Section 4 offers an alternative, but
much faster, way of deriving a continuous power spectrum estimate
between regimes (I) and (II), together with an accurate covariance
matrix. Following Section 4,  compute  two PCL
estimate $\hat C^{pw_1}_\ell$, and $\hat C^{pw_2}_\ell$, where the superscripts
denote two different weight functions. The two PCL estimates can be combined
with QML estimates at low multipoles to form a single data vector $\hat C_{\alpha \ell}$,
where $\alpha \equiv q, pw_1, pw_2$. Equations (\ref{H4}) and (\ref{H5}) can then
be solved to give a hybrid estimate $\hat C^h_\ell$ and its associated covariance
matrix. This requires the cross-covariance of the two PCL estimates, which is
straightforward to derive following the analysis of Sections 2.2 and 5.1;
simply replace the terms $w_i^2$ in equations (\ref{V2b}), (\ref{PCLN6}) and
(\ref{PCLN6b}) by the product $(w_1)_i (w_2)_i$.

If the weight function $(w_1)_i$ is chosen to be equal weight per
pixel, and weight function $(w_2)_i$ is chosen to be $(w_2)_i=
1/(\sigma_i^2 \Omega_i)$, (or a suitably regularized weighting, see
Section 5.4) then the hybrid estimate $\hat C^h_\ell$ will return
answers that are a close approximation to the ML solution {\it over
the full range of multipoles}. This method can be applied to any
number of PCL estimates, for example, one could include a third PCL
estimate with a heuristic intermediate weighting as in equation
(\ref{HN1}). In the examples described in Section 5.4 only two PCL
estimates (equal weighting and a regularized inverse variance
weighting) are used even though the the results would improve slightly
by including an intermediate weighting scheme.

 Our final procedure for estimating a power spectrum is therefore as follows:

\smallskip

\noindent
(i) Compute a  QML estimate from a low resolution map at low multipoles. If necessary,
this estimate can include a general noise matrix $N_{ij}$.

\smallskip

\noindent
(ii) Compute PCL estimates from a high resolution map using a number of
weighting schemes, {\it e.g.} equal weight per pixel and inverse variance
weighting.

\smallskip

\noindent
(iii) Combine the estimates from steps (i) and (ii) together with
their covariances and cross-covariances and solve for the hybrid
estimate $\hat C^h_\ell$ and its covariance matrix as described in
Section 4.

\subsection{Numerical examples}

\begin{figure*}

\vskip 3.0 truein

\includegraphics{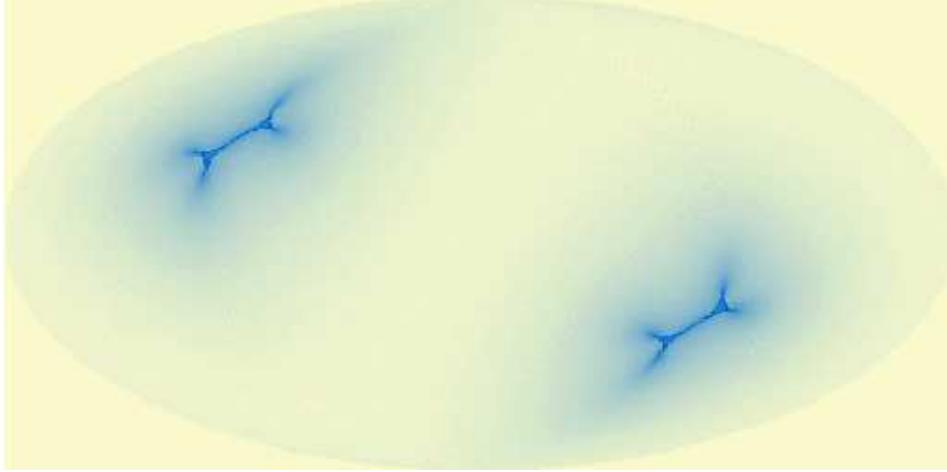}

\caption
{Map showing the hit count distribution for a Planck-like scanning strategy in
Galactic coordinates (see text for details). Darker regions close to
the ecliptic poles have a higher hit count.}

\label{figure11}

\end{figure*}

To illustrate the hybrid estimator in the presence of instrumental
noise, we have simulated the hit count distribution for a Planck-type
scanning strategy (Figure \ref{figure11}).  In this example, the
spin-axis of the spacecraft is aligned along the ecliptic plane, but a
slow precession of $\theta = 5^\circ \sin(2\phi_e)$ is applied as the
spacecraft scans in ecliptic longitude $\phi_e$. Figure \ref{figure11}
was constructed for a single detector, pointing at an angle of
$85^\circ$ with respect to the spin-axis, and for a uniform scan rate
in $\phi_e$. With this scanning geometry, the entire sky is scanned as
$\phi_e$ varies from $0$ to $2 \pi$ and the regions with high hit-count
are concentrated at the ecliptic poles as shown in Figure \ref{figure11}.
If the mean hit-count $\langle N_{\rm obs} \rangle$ is normalised to
unity, the hit-count distribution in Figure \ref{figure11} varies from a 
minimum of $0.54$ to a maximum of $120$.

\begin{figure*}

\vskip 3.2 truein

\includegraphics{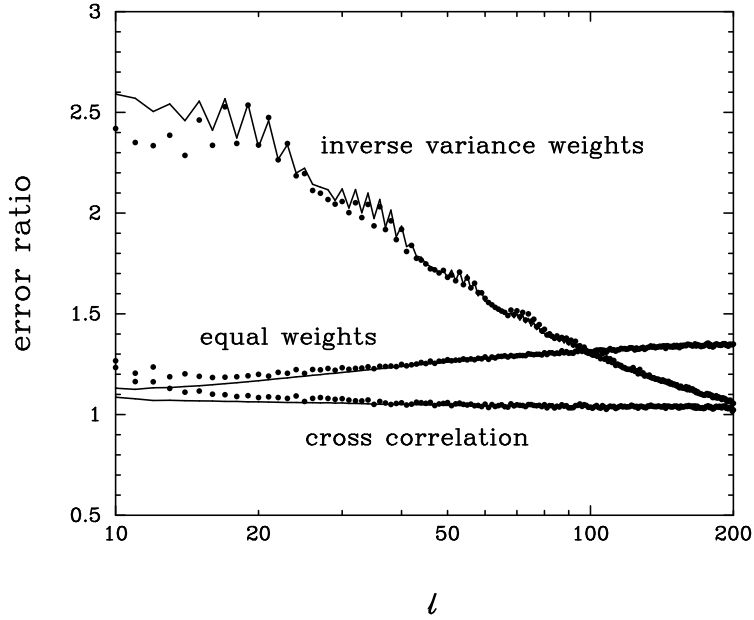}

\caption
{The points show the square roots of the diagonal components of the
covariance and cross-covariance, derived from $5 \times 10^4$
simulations, of the deconvolved PCL estimates, divided by the simple
formula of equation (\ref{NE1}). The simulations have the same
parameters as the high resolution map plotted in Figure \ref{figure7}
but include a uncorrelated noise for a Planck-type scanning  pattern as
shown in Figure \ref{figure11} (see text for details). The results are
shown for equal weight per pixel and for inverse variance
weighting. The lines show the analytic model of equations
(\ref{PCLN4}) and (\ref{PCLN7}).}

\label{figure12}

\end{figure*}

\begin{figure*}
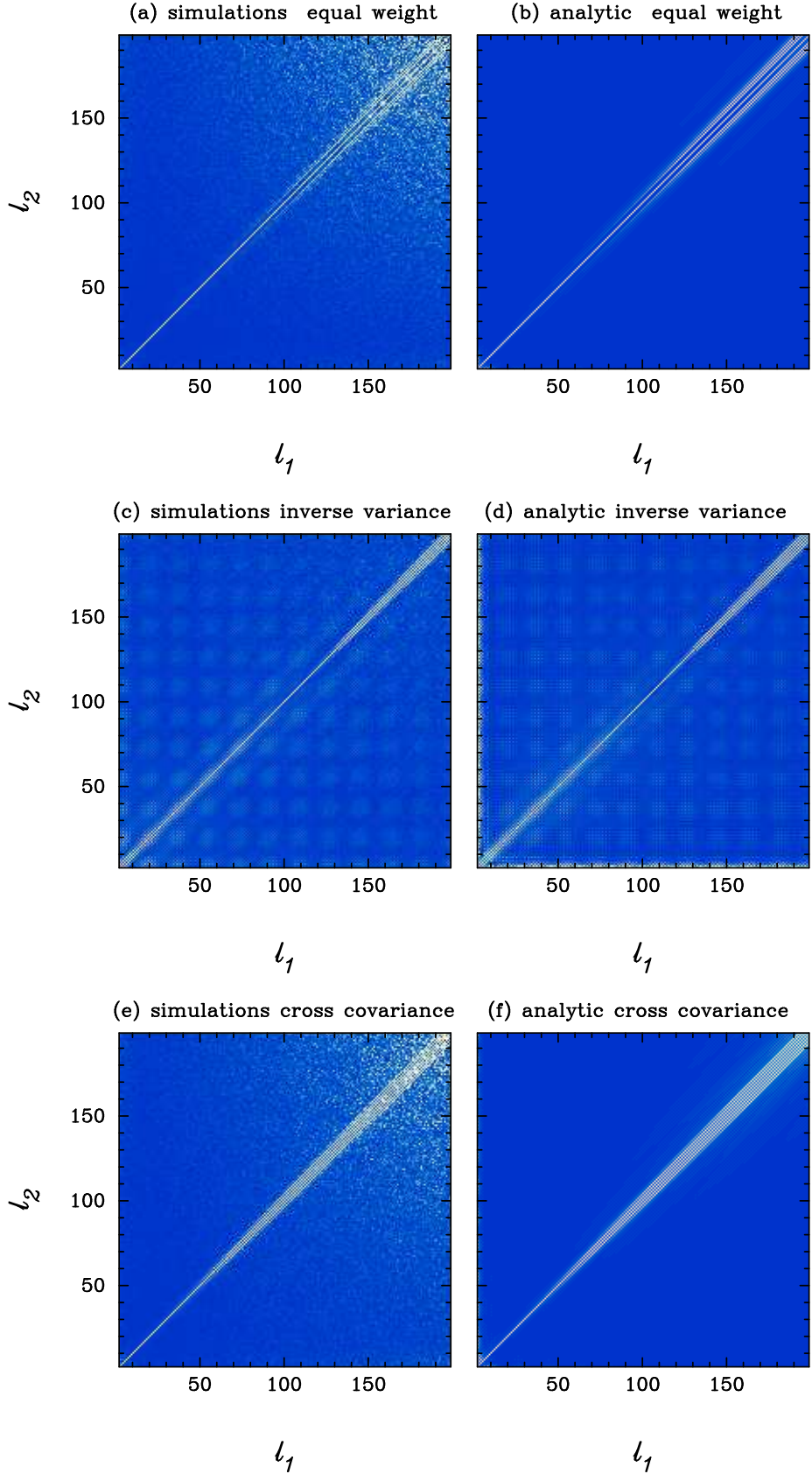


\vskip 8.8 truein

\includegraphics{pgcov_noise1.ps}
\includegraphics{pgcov_noise2.ps}

\includegraphics{pgcov_noise3.ps}

\caption
{The analytic covariance matrices for the deconvolved PCL estimates $\hat C^p_\ell$
compared with the results from
$5\times 10^4$ simulations including noise. The upper panel (figures 13a and
13b) shows the covariance matrices for equal weight per pixel. The middle
panel (figures 13c and 13d) shows the covariance matrices for inverse variance
weighting. The lower panel (figures 13e and 13f) shows the cross-covariance
between the estimates for these two weighting schemes.}
\label{figure13}

\end{figure*}

Figure \ref{figure12} compares the diagonal components of the
covariance and cross-covariance matrices for deconvolved PCL
estimates, $\hat C^p_\ell$, for a set of $5 \times 10^4$ simulations
against the analytic formulae given in Section 5.1. The simulations
have the same parameters as the high resolution map shown in Figure
\ref{figure7} ($\theta_c=0.5^\circ$, $\theta_s=1^\circ$) and the noise
level was normalized to $\sigma_p = 1 \times 10^{-4}$ for pixels with
the mean hit count. As in previous Sections a Galactic cut of
$\pm 10^\circ$ was imposed. The filled circles in Figure
\ref{figure12} shows the square roots of the diagonal components of
the covariance and cross-covariance matrices divided by the
approximate formula (Knox 1995)
\begin{equation}
\langle  (\Delta \hat C_\ell)^2 \rangle^{1/2} \approx \left ( {2 \over (2 \ell + 1) f_{\rm sky}}
\right )^{1/2}
( C_\ell + \sigma^2_p \Omega_p). \label{NE1}
\end{equation}
As expected from the analysis of Section 5.1, inverse noise
weighting leads to smaller errors than equal weighting at $\ell \simgt
100$ when the noise dominates the error budget.  However, inverse
variance weighting leads to much poorer estimates of the power
spectrum at $\ell \simlt 100$. This is caused by the cusp-like
structure of the hit count distribution shown in Figure \ref{figure11}. For
this scanning pattern, the small number of pixels with a high hit
counts close to the ecliptic poles dominate the PCL estimates with
inverse variance weighting, consequently the errors on the
PCL estimates at low multipoles are much larger than expected from
cosmic variance. This scanning strategy was chosen intentionally to
illustrate the differences between inverse variance weights and equal
weights. The differences between these weighting schemes would be
smaller for a WMAP-like scanning strategy, which produces a smooth
variation of the hit counts with ecliptic latitude.

\begin{figure*}
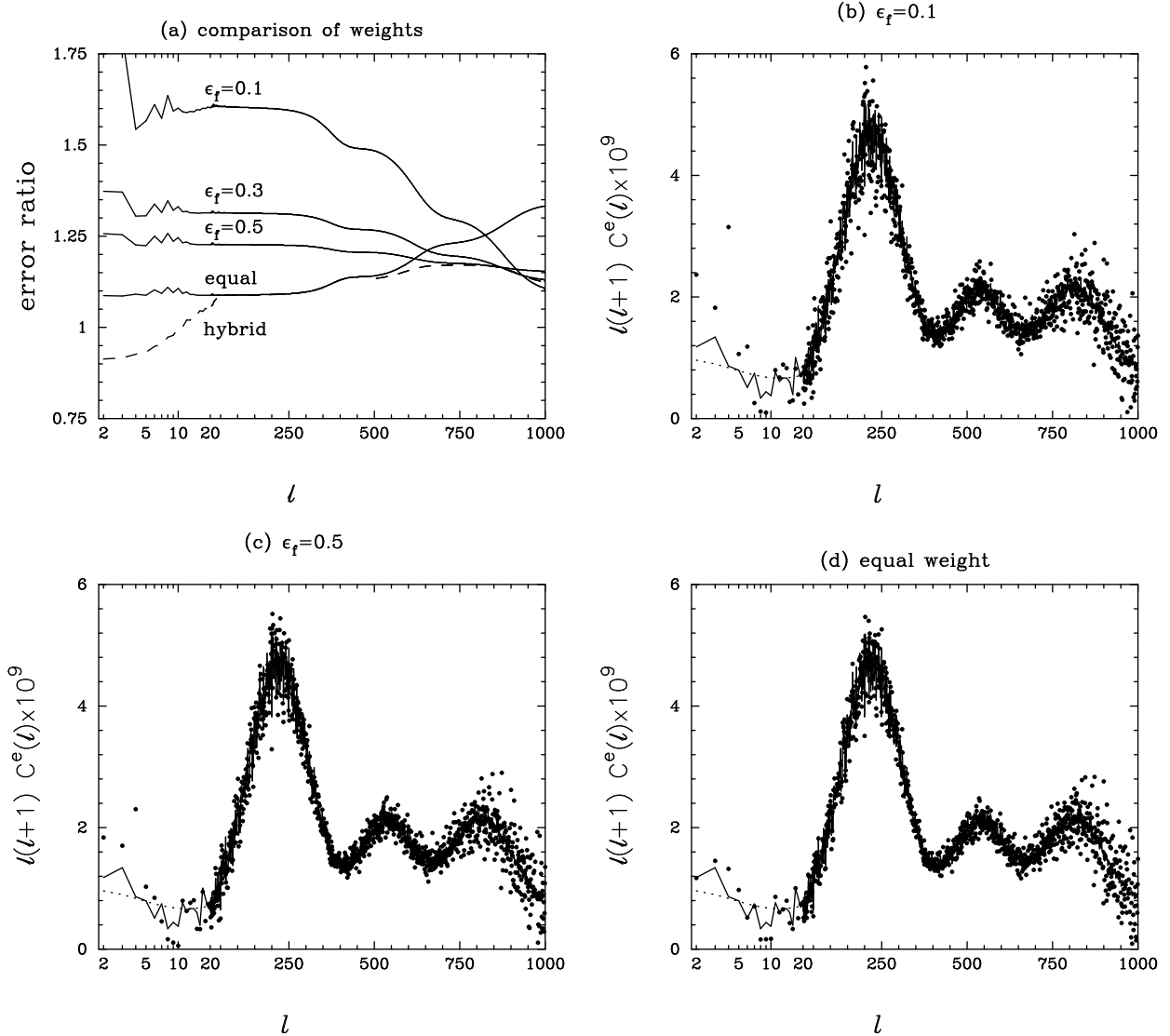


\vskip 5.8 truein

\includegraphics{pgmaster_err_1000.ps}
\includegraphics{pgmaster_1000b.ps}

\includegraphics{pgmaster_1000c.ps}
\includegraphics{pgmaster_1000d.ps}

\caption
{Figure 14a shows the analytic dispersions, normalized by equation
(\ref{NE1}), for various weighting schemes. These have been computed
for a map with $0.1^\circ$ pixels, a smoothing of $\theta_s =
0.2^\circ$, and a Planck-type noise pattern as shown in Figure
\ref{figure11}. The noise level was set to $\sigma_p= 2\times 10^{-5}$
for a pixel with the average hit count. The parameter $\epsilon_f$ is
the regularizing parameter in the weighting scheme of equation
(\ref{NE2}). The dashed line in Figure \ref{figure14}a was computed
from the diagonal components of the covariance matrix for the hybrid
estimator (equation \ref{H5}) applied to a data vector consisting of
QML estimates over the range $2 \le \ell \le 40$, PCL estimates with
equal weight per pixel over the range $2 \le \ell \le 1000$, and PCL
estimates with $\epsilon_f=0.5$ weighting over the range $500 \le \ell
\le 1000$. Figures (b)-(d) show the PCL estimates for the same simulation
for three different weighting schemes. The solid line shows the power
spectrum for this particular simulation and the dotted line shows the
power spectrum of the fiducial $\Lambda$CDM model. The power spectra
in these figures have been corrected to zero beam width. Note that the
abscissae in all of these plots uses a mixed logarithmic-linear scale in
$\ell$,  so that the behaviour of the power spectrum at low multipoles
is clearly visible; the scale is logarithmic over the rangle $2 \le \ell
\le 20$ and linear at higher multipoles.}
\label{figure14}

\end{figure*}

Another consequence of the cuspiness of the hit count distribution of
Figure \ref{figure11} is that the window function $\tilde W^{(2)}$
(equation (\ref{V2b}) for inverse variance weighting is broader the
window function for equal weights. For inverse variance weighting, the
analytic approximation to the covariance matrix $\tilde V_{\ell
\ell^\prime}$ is therefore not as accurate at low multipoles as it is
for equal weights. The full matrices are compared with the analytic
formulae in Figure \ref{figure13}. As expected, the analytic formulae
are in excellent agreement with the results from the simulations at
high multipoles. However, there are large discrepancies at low
multipoles in the case of inverse variance weighting that are clearly visible
in Figures \ref{figure13}c and \ref{figure13}d.  As described in
Section 2.2, the covariance matrices (and cross-covariances) at low
multipoles can be evaluated accurately by summing over pixels using
degraded resolution maps ({\it cf} equation \ref{V6}), but as we
will see in the next example, there is no need to compute these
components for the hybrid estimator.

As a final illustration of the hybrid estimator, we have analysed a
simulation with $0.1^\circ$ pixels and a beam smoothing of $\theta_s =
0.2^\circ$. The simulated map contains $4.12 \times 10^6$ pixels and is
therefore comparable in size  and resolution to the CMB maps produced
by WMAP. A Planck-like scanning strategy, as shown in Figure \ref{figure11},
was adopted and the  noise level was set to $\sigma_p = 2\times 10^{-5}$ 
for pixels with the mean hit count. At this resolution, the cuspiness
in the hit count distribution  causes numerical problems in evaluating the PCL esimator
at high multipoles if exact inverse variance weighting is used. These
can be avoided by introducing a regularising parameter $\epsilon_f$,
\begin{equation}
w_i = {\sigma^2_p \langle \Omega_i \rangle
\over (\sigma^2_i + \epsilon_f \sigma^2_p) \Omega_i}, \label{NE2}
\end{equation}
in the weight function ({\it cf} the heuristic weight function of equation
(\ref{HN1})). The analytic dispersions for the deconvolved PCL estimates
 (divided by the approximation of equation \ref{NE1}) 
are shown in Figure \ref{figure14}a for four weighting schemes.
The errors for equal weighting grow steadily at $\ell \simgt 500$, as
noise becomes a more important contribution to the errors (the 
structure in this curve at $\ell \simgt 500$,
{\it i.e.} the changes in slope and the  plateaus, are related to the
acoustic peak structure in the CMB spectrum). The weighting scheme
of equation (\ref{NE2}) with  $\epsilon_f = 0.1$ leads to 
an improvement in the errors at $\ell \simgt 800$, but performs
poorly at lower multipoles. The weighting with $\epsilon_f=0.5$
performs almost as well as the $\epsilon_f=0.1$ weighting in the 
noise dominated limit, but performs much better in the intermediate
region $500 \simlt \ell \simlt 750$. We therefore use the equal weights
and $\epsilon_f=0.5$ PCL estimates in the hybrid estimator.

\begin{figure*}

\vskip 4.2 truein

\includegraphics{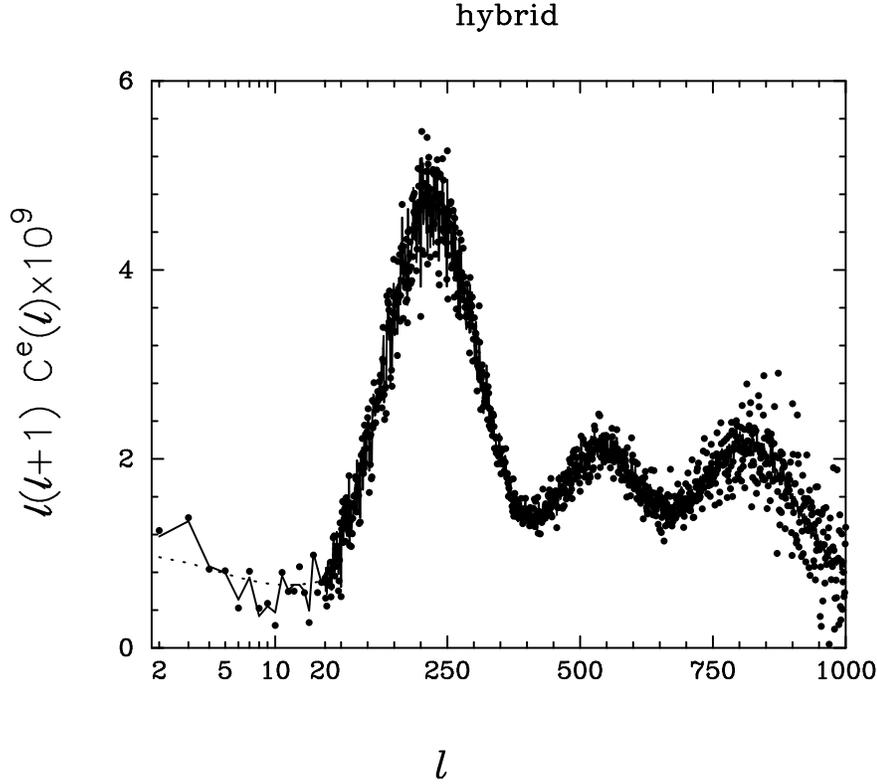}

\caption
{The filled circles show the hybrid estimate $\hat C^h_\ell$ for the
high resolution simulation, including noise, used for the PCL
estimates in Figure \ref{figure14}. As in Figure \ref{figure14} the
dotted line shows the fiducial $\Lambda$CDM power spectrum and the
solid line shows the actual power spectrum computed from the $a_{\ell
m}$ coefficients used to generate the map. The hybrid estimator was
applied to a data vector consisting of QML estimates over the range $2
\le \ell \le 40$, PCL estimates with equal weight per pixel over the
range $2 \le \ell \le 1000$, and PCL estimates with $\epsilon_f=0.5$
weighting over the range $500 \le \ell \le 1000$. The hybrid estimator
therefore returns almost the exact input power spectrum up to
$\ell=40$, smoothly matches on to the equal variance PCL estimates up to
$\ell \sim 700$ and then smoothly matches to the $\epsilon_f = 0.5$ PCL
estimates at higher multipoles.}

\label{figure15}

\end{figure*}

 The inputs into the hybrid estimator are as follows:

\smallskip

\noindent
(i) QML estimates and Fisher matrix for $\ell \le 40$ computed from a degraded
resolution map (pixel size of $\theta_c=3^\circ$, smoothing $\theta_s=3^\circ$).

\smallskip

\noindent
(ii) PCL estimate with equal weight per pixel over the range $2 \le \ell \le 1000$
with analytic covariance matrix (equation \ref{PCLN4}) for multipoles $\ell \ge 30$.
At lower multipoles, the covariance matrix was computed by summing over pixels
(equation (\ref{V6})) in a lower resolution map ($\theta_c = 1.5^\circ$, $\theta_s=1.5^\circ$).

\smallskip

\noindent
(iii) PCL estimates using weights with $\epsilon_f = 0.5$ (equation \ref{NE2})
over the range $500 \le \ell \le 1000$ together with the analytic covariance matrix.

\smallskip

\noindent
(iv) The cross-covariances between equal weight PCL estimates and QML estimates 
(equation \ref{H1}).

\smallskip
\noindent
(v) The cross-covariances between equal weight and $\epsilon_f=0.5$ PCL estimates, 
computed from equation (\ref{PCLN4}) with elements at $(\ell, \ell^\prime) \le 100$
set to zero to avoid any spurious coupling from inaccuracies in the analytic
expression at low multipoles.

\smallskip

The hybrid estimates are plotted in Figure \ref{figure15} using the
same logarithmic-linear abscissa as in Figure \ref{figure14}. At low
multipoles, the power spectrum estimates reproduce almost exactly the
input power spectrum for this simulation (solid line) as in the
noise-free example shown in Figure \ref{figure9}. The estimates then
match smoothly to the equal weight PCL estimates (Figure
\ref{figure14}d) at $\ell \ge 40$, and then match smoothly on to the
$\epsilon_f = 0.5$ PCL estimates (Figure \ref{figure14}c) at $\ell
\simgt 700$. The dispersions of these estimates computed from equation
(\ref{H5}) are plotted as the dashed line in Figure
\ref{figure14}b. This figure shows that the hybrid estimates are close
to optimal over the full range of multipoles $2 \le \ell \le 1000$.
There would be some small improvement in extending the QML estimator to
higher multipoles, since the QML errors have not quite converged to
the PCL errors by $\ell =40$. It would also be possible improve the
errors over the multipole range $500 \simlt \ell \simlt 700$ by adding
a third PCL estimate with $\epsilon_f \sim 1$ into the hybrid
estimator, but any improvement will be small. The covariance matrix
for the hybrid estimator is almost exactly diagonal at low multipoles,
as in the example shown in Figure \ref{figure10}, and becomes band
diagonal at higher multipoles, as in the examples shown in Figures
\ref{figure13}a and \ref{figure13}c.

\section{Conclusions}

As explained in the Introduction, a number of methods of estimating
power spectra have been discussed over the last few years. These range
from the fast PCL estimators, first applied to cosmology in the
1960's, to more sophisticated ${\cal O}(N_d^3)$ ML estimators. The
Introduction reviewed these methods and contrasted their strengths and
weaknesses.  We concluded that for the analysis of realistic data,
what is required is a {\it fast} power spectrum estimator that returns
a reliable covariance matrix.  Such an estimator can then be used as
part of a Monte Carlo chain to assess the effects of various complex
sources of errors, unavoidable in real experiments, on the power
spectrum estimates and their covariances. This point of view is
consistent with the MASTER\footnote{Monte Carlo Apodised Spherical
Transform Estimator} method of Hivon \etals (2002) and with the
analysis of the WMAP data described by Hinshaw \etals (2003), both of
which are based on fast PCL estimators.

In this  paper, an unbiased hybrid estimator was developed that
combines QML estimates at low multipoles and PCL estimates with
various weightings at higher multipoles. A number of analytic results
were derived showing the statistical equivalence of QCL and PCL
estimators in the noise-free and noise-dominated limits. Expressions
for the covariance matrices and their cross-covariances were derived
and tested against numerical simulations. The hybrid estimator was
illustrated for a high resolution CMB experiment ($\theta_s =
0.2^\circ$) and shown to reproduce the power spectrum with close to
optimal errors over the full range of multipoles $2 \le \ell \le 1000$
for a Planck-type noise pattern. The method is fast (the timing is
dominated by the ${\cal O}(N_d)^{3/2}$ operation count for fast
spherical transforms) and returns an accurate and nearly band-diagonal
covariance matrix over the full range of multipoles. The hybrid
estimator therefore fulfills the criteria discussed in the previous
paragraph for a fast power spectrum estimator, with a simple covariance matrix,
 suitable for inclusion in a Monte Carlo chain.

In the numerical examples described in this paper, the instrumental
noise was assumed to be diagonal. This is, of course, unrealistic and
is the main limitation of the analysis presented in this paper. The
instrumental (and other) sources of noise will have $1/f$-type noise
properties. When coupled with a Planck-type scanning strategy,
$1/f$-type instrumental noise will lead to correlated noise on the sky
(in the form of stripes in the sky maps). The effects of striping can
be reduced during the map-making stage, either by applying an optimal
maximum likelihood map-making algorithm ({\it e.g.} Wright 1996;
Dor\'e \etal, 2001b, Natoli \etal, 2001) or by applying a simpler but
sub-optimal destriping algorithm ({\it e.g.} Delabrouille J, 1998;
Keih\"anen \etal, 2003).  The QML estimates can, of course, handle
correlated pixel noise described by the general noise matrix $N_{ij}$
of equation (\ref{QN1}). Thus a simple way of accounting for $1/f$
detecter noise on the low CMB multipoles is to construct a degraded
resolution map directly from the TOD, together with the full noise
covariance matrix for the degraded resolution maps. It is
computationally feasible to produce such low resolution maps ($\simlt
10^4$ pixels) and covariance matrices using maximum likelihood map
making methods from Planck-sized TODs (Borrill, private
communication). The low resolution data, which need only be computed
once, can then be fed as inputs into the QML estimator as described in
Section 5.

This type of analysis could eliminate the requirement in the MASTER
method for the calibration of power spectrum transfer functions via
simulations (see equation 15 of Hivon \etals 2002). In addition, the
covariance matrix of the hybrid estimator would reflect the effects of
correlated noise at low multipoles accurately, since it uses the QML
Fisher matrix as an input. However, in this approach, the effects of
$1/f$ noise would not be properly included in the PCL
estimates. Fortunately, for the parameters of the Planck instruments,
the effects of $1/f$ noise should be confined to low multipoles ($\ell
\simlt 100$, see {\it e.g.} Keih\"anen \etals 2003) and the
pixel-pixel noise should be strongly diagonally dominated ({\it e.g.}
Stompor and White 2003) suggesting that PCL estimates and the analytic
estimates of their covariance matrix should be accurate. The effects of
realistic $1/f$ noise on the hybrid estimator clearly need to tested
against numerical simulations.

 In the examples shown in this paper, the QML estimator was applied to
low resolution maps with a maximum of $3842$ pixels. However, it is
possible to analyse larger maps within a Monte Carlo chain. For any
particular sky cut and noise covariance matrix, the matrices $E_{ij}$,
the Fisher matrix and the cross covariances with PCL estimates need
only be computed once and stored. The evaluation of the QML estimates
for different realisations of the data vector $x_i$ then requires only
the evaluation of traces (equation \ref{QN1}), provided that the
matrices can be read into the computer memory. The main limitation is
likely to be set by the size of the computer memory, rather than the
cpu time available. This time-saving trick was used to evaluate the
QML estimates from the large number of simulations used to construct
Figure \ref{figure5}.

Throughout this paper we have assumed spherically symmetric beams and
that the beam profiles are known exactly. Neither of these assumptions
will be true in practice. Uncertainties in the beam shapes would
affect the PCL estimates at high multipoles. Such uncertainties can be
incorporated in the PCL covariance matrix as described by Hinshaw
\etals (2003). Dealing with far-field beam asymmetries is more
problematic as a calibration of a PCL estimator would require
simulations of the scanning strategy with a full $4 \pi$ convolution
of the beam pattern on the sky (Wandelt and G\'orski 2001). For a
mission such as Planck, the beam asymmetries are small enough that an analysis
of the CMB power spectrum  in terms of an effective spherically symmetric beam should
be adequate. As with $1/f$ noise, the effects of beam asymmetries on the
Planck mission need to be checked against  numerical simulations.

The principal mythology that this paper is designed to dispel is that
a full ${\cal O}(N_d^3)$ maximum likelihood solution will produce some
magical improvement in estimates of a power spectrum. The truth is
that for realistic experiments, any benefits over the hybrid estimator
are likely to be small ({\it cf} Figure \ref{figure15}) and
overwhelmed by `real word' uncertainties such as beam calibrations,
foreground separation, point source subtraction {\it etc}. The hybrid
estimator leads to an unbiased, nearly optimal, power spectrum
estimator with an accurate and almost diagonal covariance matrix. The
benefits of being able to assess such real world complexities using a
hybrid estimator within a Monte Carlo chain, and hence to quantify any
corrections to the covariance matrix, are likely to more than offset
any marginal improvements that might be gained from a full ${\cal
O}(N_d)^3$ ML solution. Since the hybrid estimator is fast, and the
covariance matrix is nearly diagonal, it is possible to test simple
parametric scalings of the covariance matrix to produce a good
approximation to the $\chi^2$ of equation (\ref{NR4}) for different
theoretical power spectra. It is also possible to test
parameterizations of the shape of the likelihood function, rather than
using the $\chi^2$ model of equation (\ref{NR4}). (This should not
require large numbers of simulations if the covariance matrix of the
hybrid estimator is diagonally dominant as
in the examples discussed in Section 5.)  Both of these tests are
important for the unbiased estimation of cosmological parameters (see
{\it e.g.} Verde \etals 2003) and will become of still greater
importance for the analysis of high precision experiments such as
Planck.

The hybrid estimator can be generalised to handle other types of data,
for example, galaxy clustering, weak gravitational lensing and CMB
polarization. The extension to CMB polarization will be discussed in a future
paper.

\medskip

\noindent
{\bf Acknowledgements:} I thank members of the Cambridge Planck Analysis Centre,
especially Anthony Challinor, for helpful discussions. I thank the referee for
useful comments.

\end{document}